\documentclass[pdflatex,sn-mathphys]{sn-jnl}

\jyear{2022}%

\theoremstyle{thmstyleone}%
\newtheorem{theorem}{Theorem}
\theoremstyle{thmstyletwo}%
\usepackage{nicefrac}
\theoremstyle{thmstylethree}%
\newtheorem{definition}{Definition}%
\newcommand{\ignore}[1]{}
\newtheorem{thm}{{\bf Theorem}}
\newtheorem{cor}{{\bf Corollary}}
\newtheorem{lemma}{{\bf Lemma}}
\newcommand{\eop}{{\hfill $\blacksquare$} }
\newcommand{\Sw}{{\cal S}} 
\newcommand{\Nw}{{\cal N}} 
\newcommand{\Dw}{{\cal D}} 
\newcommand{\up}{\Theta}

\newcommand{\Ups}{\Theta}
\newcommand{\ups}{\Theta}

\usepackage{multirow}
\begin{document}

\title[Systemic-risk and  evolutionary stable strategies in a financial network ]{Systemic-risk and  evolutionary stable strategies in a financial network}   


\author[1]{\fnm{Indrajit} \sur{Saha}}\email{indrojit@iitb.ac.in}
\author*[2]{\fnm{Veeraruna} \sur{Kavitha}} \email{indrojit@iitb.ac.in}\email{vkavitha@iitb.ac.in}

\affil*[1]{\orgdiv{Industrial Engineering and Operations Research}, \orgname{Indian Institute of Technology Bombay}, \orgaddress{\street{Powai}, \city{Mumbai}, \postcode{400076}, \state{Maharashtra}, \country{India}}}

\abstract{

We  consider a   financial network represented at any time instance by a random liability graph  which evolves over time. The agents connect through credit instruments borrowed from each other or through direct lending, and these create the liability edges. These random edges are modified (locally)  by the agents over time, as they  learn from their experiences and (possibly imperfect) observations. The settlement of the liabilities  of various agents at the end of the contract period (at any time instance) can be expressed as solutions of random fixed point equations. Our first step is to derive the solutions of these equations (asymptotically and one for each time instance), using a recent result on random fixed point equations.
The agents, at any time instance, adapt one of the two available strategies, risky or less risky investments, with an aim to maximize their returns. 
We aim to study the emerging strategies of such  replicator dynamics that drives the financial network.  We theoretically reduce the analysis of the complex system to that of an appropriate ordinary differential equation (ODE). 
Using the attractors of the resulting ODE we show that the replicator dynamics converges to one of the two pure evolutionary stable strategies (all risky or all less risky agents);  one can have mixed limit only when the observations are imperfect.  
We   verify our theoretical findings using exhaustive Monte Carlo simulations.
The dynamics avoid the emergence of the systemic-risk regime (where majority default).  However, if all the agents blindly adapt risky strategy it can lead to
systemic risk regime.
}


\keywords{Evolutionary  stable strategy,   Replicator dynamics, Ordinary differential equation, Systemic risk, Financial network.}

\pacs[JEL Classification]{C73, G11}


\maketitle

\section{Introduction}
\label{Sec_Intro}
We consider a   financial network,  where the agents are interconnected  to each other through financial  commitments  (e.g., borrowing, lending, etc).  In addition they invest in either risk-free (bonds, fixed deposit, government projects etc) or risky securities (derivative markets, stocks etc). In such a system the agents  not only face  random economic shocks (received via significantly smaller returns of their risky investments), they are also affected by the percolation of  the shocks faced by their neighbours (creditors/borrowers), neighbours of their  neighbours and so on. 
 In the recent years from $2007$-$2008$ onwards,  there is a surge of activity to study the financial and systemic level risks  caused by such a percolation of shocks (\cite{acemoglu2015systemic,allen2000financial,eisenberg2001systemic,kavitha2018random}).
Systemic risk is the study of the risks related to  financial networks, when individual or entity level shocks can trigger   severe instability at system level that can collapse the entire economy (e.g.,   \cite{acemoglu2015systemic,allen2000financial,eisenberg2001systemic}). In this set of papers, the authors study the kind of topology (or graph structure)  that is more stable towards the percolation of shocks in a financial network, where stability is measured in terms of     number/fraction of defaults.  

In contrast to many existing studies   related to systemic risk, we consider heterogeneous agents and an evolutionary framework. There are two  groups of agents   existing simultaneously  in the  network;  one group of agents invest in  risk-free assets,  while the   rest consider risky investments. The second group borrows   money from   other agents of the network to gather more funds towards  the risky investments  (with much higher expected returns). 
These investments  are subjected to large (but rare) economic shocks,   
which can potentially percolate  throughout  the  network and can also affect the `less risky' agents. The extent of percolation depends  upon   the relative sizes of the two groups.

We further consider that the system evolves over time, some strategic agents change their choices (investment types) and we also have agents randomly joining the network at various time instances. 
  The new agents choose their investment type (risky or less risky) based on  their observations of the  returns   of a random sample of agents that invested  in the  previous round. The existing agents may change their strategy based on their own returns and their observations of   a random sample. 
   The relative sizes of the two groups changes, the network structure changes,  which influences the (economic shock-affected)  returns  of the agents in the next round, which in turn influences the decision of the  agents for the round after. Thus the system evolves after each round. We study this   process  using the well known evolutionary game theoretic tools.

 From the perspective of a financial network,  this type of work is new to the  best of our knowledge. We found few papers that consider evolutionary approach in  other aspects related to finance; in \cite{yang2016evolutionary},  the authors study the financial safety net (a  series of the arrangement of the firms to maintain financial stability), and  analyze the   evolution of the bank strategies  (to take insurance or not); 
recently in  \cite{LiHonggang} authors consider an evolutionary game theoretic model with three types of players, i) momentum traders ii) contrarian traders iii) fundamentalists and studied the evolution of the relative populations. As already mentioned, these papers relate to very different aspects in comparison with our work.

 In  \cite{Saha} we  have an initial analysis of this flavor related to systemic risk analysis; the new agents choose their strategy once and for all when they join the network; we showed  that all the agents eventually revert to either risky or   less risky portfolios unless they choose their strategies based on a large sample of observations. In the former scenario, the dynamics converge to a pure (only one of the strategies, risky/less risky, survive) evolutionary stable strategy (ESS), while the latter scenario converges to a mixed ESS. In \cite{Saha}, we  derive the analysis, only for the cases in which the  dynamics are smooth and always converge to one solution (ESS) for a given set of parameters. 
  
The current paper considers a significant generalization.
We consider agent  modifying their strategies at various time instances, in addition to the agents that randomly join the network and choose their investment type.  
These types of processes are often analyzed using ordinary differential equations (ODEs).  
The resulting ODEs in this context are non-smooth and we completed the analysis of such a complex ODE, by using the approximate solutions of   a relevant set of random fixed point equations (derived using the recent results of \cite{kavitha2018random,saha2021random}).  The analysis of the simplified representations of the fixed points helped us in deriving the attractors and the corresponding regions of attraction of the otherwise complex (non-smooth) ODE. 
We finally showed that the network can converge to one of the two pure ESSs based on network parameters. More importantly, this convergence also depends upon  the initial configuration.  
Further,  the dynamics  avert the systemic risk regime by converging to the pure ESS of all less risky agents, in scenarios that have the potential to lead to the systemic risk event (when the majority of the agents fail to clear their obligations). One may have mixed ESS when the observations are partially erroneous. Interestingly even these imperfect observations are (majorly) sufficient to drive the system away from the systemic risk regime.

Our analysis can also be used to show that the system actually leads to systemic risk regime, when users blindly adapt risky strategy; this is true  in scenarios where the rational agents have avoided it, i.e., when the ESS is not equal to all risky agents. In other words, when ESS is pure and has all risky agents, then there is no difference between rational (those that observe and change strategies) and blind (adapting only risky strategy) agents; all risky agents is ESS, only when the probability of default is zero; in the remaining scenarios the rational agents would avoid systemic risk regime, while, with blind agents the system can lead to a situation where all the agents default.


The analysis of these complex networks  (in each round) necessitated  the study of  random fixed point equations (defined sample path-wise in large dimensional spaces), which represent the clearing vectors of all the agents (\cite{acemoglu2015systemic,eisenberg2001systemic} etc).  As already mentioned, the study is made possible because of the recent result in \cite{kavitha2018random,saha2021random}, which  provided an asymptotically accurate  one-dimensional  equivalent solution.

We conducted an exhaustive Monte Carlo (MC) simulation-based numerical study with the following details: a) the dynamics (economic shocks, random returns, random set of agents modifying their choices,  etc.) are randomly generated; and b) the random clearing vector is obtained by numerically solving the relevant fixed point equations, in each round; however, we did not generate random liability connections in each round.  The limits of  these dynamics well match with the attractors of the corresponding ODE. 

Using the random fixed point approximation, we also derive the closed-form expressions for the solution of the ODEs approximating the financial dynamics. Once again using exhaustive MC simulations we showed that the ODE trajectories well approximate  the random financial trajectories, even for a finite (initial) number of rounds. The approximation is good even for a few hundred of entities and even for a few initial ($\approx$ 100) rounds.  
 
\textbf{Organization:} The financial network and random fixed point solutions for clearing vectors are discussed respectively in sections \ref{sec_model} and \ref{sec_aymptotic}. The ODE approximation is derived in Section \ref{Sec_finance_replicatordynamics}, while (mixed) evolutionary stable  strategies are discussed in Section \ref{sec_ess}.  Section \ref{Section_defaultersleave} considers the system when some  defaulted agents may stop investing in further rounds.
Numerical examples are  provided in Section \ref{Section_numerical_observations}.

 \section{ Finance Network}
\label{sec_model}

We consider 
random graphs, where the edges represent the financial connection between the  nodes. Any two nodes are connected with probability $p_{ss} > 0$ independent of the others.
This graph represents a   financial network where borrowing and lending are represented by   edges and the weights over them. 
The modeller may not have access to  exact  connections of the network,  but a random graph  is a good model to analyse such  complex systems. In particular we consider   graphs that satisfy the assumptions of \cite{kavitha2018random,saha2021random}.  
  
\noindent The agents are repeatedly investing in some financial projects. In each round of investment, the   agents borrow/lend   from/to some   random subset of the agents of the network. Some of them may invest the remaining
  in a risk-free investment (which has a constant rate of interest $r_s$).  While the others invest the rest of their money in risky investments which have random returns; we consider a binomial model in which returns are high (rate $u$) with high probability $\delta$ and can have large shocks (rate $d$), but with small probability ($1-\delta$); it is clear that $d < r_s < u$.  We thus have two types of agents, we call the group that invests in risk-free  projects as `less risky'  group  ($G_1$), the rest are   referred to as `risky' group ($G_2$).
  
 New agents may join the network in each round of investment. They  choose their investment type, either less risky or risky,  for the first time based on the previous  experience of the network.   
  The new agents learn from the returns of the agents of the previous round of investments before choosing  a suitable investment type.
  To be precise, they   learn from the experience of a  random sample (returns of  two random agents) of the network corresponding to the previous round. 

  In any round, 
  a random agent of the network may modify its  strategy,   based on its own  experience and  that of another random agent of the network. 

{\bf Two strategies:} As mentioned before, there are two strategies available in the financial market. Less risky agents  of $G_1$ use strategy 1;  these agents lend some amount of their initial wealth to other agents (of $G_2$) that are willing to borrow,  while the rest is invested in government security, for example, bonds, government project, etc.  Risky agents of $G_2$  are  adapting strategy $2$, wherein they borrow funds from the other agents and invest in risky security, for example, derivative markets, stocks, corporate loans, etc.
These agents also lend to other agents of $G_2.$
Let  $\epsilon_t $ be the  fraction of the agents in  $G_1$ group and let $n(t)$ be the total number of agents in round $t$.   Thus the total number of agents (during round $t$) in group 1 equals  $n_1(t) :=  \lvert G_1\rvert   =  n(t) \epsilon_t$, similarly $n_2(t) := \lvert G_2 \rvert   =  n(t) (1 -\epsilon_t )$.

We consider that some new agents may join in each round and a random number of the existing agents may modify their choice. 
Thus the size of the network is increasing. The agents are homogeneous, i.e., they reserve  the same wealth $w >0$ for investments (at the  initial investment  period) of each round. 
{\it Each round is composed of two time periods, the agents invest during the \underline{initial investment period} and they obtain their returns after some given time gap.} 
The two time period model is borrowed from  \cite{acemoglu2015systemic,eisenberg2001systemic,kavitha2018random} etc.
 Recall the agents make their choice for the next round,  based on their observations of  these returns. 

{\bf Initial investment phases:}   During the initial investment phases (of any round $t$), any agent    $i\in G_{1}$  lends  to any agent $j \in G_2$ with probability $p_{ss}$ and it lends equal amount $w/n(t) p_{ss}$ to each of the approachers  that approached it for loan;  let $I_{ij}$ be the indicator of this lending event.  Thus any agent of $G_1$ lends approximately  $w (1-\epsilon_t)$ fraction  to agents of $G_2$;  note that for large $n(t)$, the number  of approachers of $G_2$ approximately equals $n(t) (1-\epsilon_t) p_{ss}$. The agents of $G_1$  invest the rest $w\epsilon_t$ in risk-free investment (returns at fixed interest-rate~$r_s$). %
We now provide complete details of the network in one round, when $\epsilon_t = \epsilon.$

Let   $\tilde{w}$ be the amount  accumulated  by any agent of  $G_2$ out of which a positive fraction $\alpha$  is invested towards the other banks of $G_2$ and $(1-\alpha)$ portion is invested in risky security. These amounts could be random and different from agent to agent, but with large networks (by the law of large numbers) one can approximate  these to be constants.
\noindent Thus   the amount accumulated by a typical $G_2$ agent is given by the following: \vspace{-2mm}
\begin{equation}
\tilde{w}  =\hspace{-2mm} \underbrace{w+ w\epsilon }_\text{Initial wealth + Borrowed from $G_1$} \hspace{-2mm}+\underbrace{\tilde{w} \alpha,}_\text{Borrow from $G_2$}  \mbox{ and hence } 
\end{equation}
$\tilde{w} = \frac{w(1+\epsilon)}{ (1-\alpha)}$; this amount is used for investments/lending in the initial phase.
Once again this equation is accurate when $n(t)$ is large and is used   for asymptotic analysis.
 Thus the total  investment towards the risky venture equals  $\tilde{w} (1-\alpha)= w(1+\epsilon)$.
The $G_2$ agents have to settle their liabilities at the end of the return/contract period (in each round) and  this would depend upon their returns from the risky investments.   Thus the total liability of any agent of  $G_2$ is  $y = (w\epsilon +\tilde{w}\alpha )(1+r_b)$, where $r_b$ is  the borrowing rate; by  simplifying 
\begin{equation}
\label{Eqn_liability}
y= \frac{w(\epsilon +\alpha)(1+r_b)}{( 1-\alpha)}.
\end{equation}
For simplicity of explanation, we are considering constant terms to represent all these quantities, in reality, they would be i.i.d. quantities which are further independent of other rounds, and the asymptotic analysis would go through as in \cite{kavitha2018random,saha2021random}.
Similarly, any agent of $G_2$ lends the following amount to each of its approachers (of $G_2$):\vspace{-3mm}
\begin{equation}
\label{Eqn_liabOne}
\frac{\alpha {\tilde w} }{n(t) (1-\epsilon_t) p_{ss}} = \frac{ \alpha w(1+\epsilon)}{n(t) (1-\epsilon_t) p_{ss}  (1-\alpha)} .
\end{equation}

{\bf Returns and Clearing Vectors:}
\ignore{ The banks have heterogeneous believe that the future value of the asset price will go up.  Therefore some of the banks are borrowing more and invest the money for a higher return while the other group of banks make a less risky investment. Although the channel of risk for the $G_{1}$ banks is inherent because the failure of the group $G_{2}$ banks can cause the portfolio value of the banks goes down.
}
We fix round $t$ and avoid notation $t$ for simpler representation. The agents of $G_2$ have to clear their liabilities during this phase in every round.  
Recall the agents of $G_2$ invested $w(1+\epsilon)$ amount in risky-investments and the corresponding random returns (after economic shocks) are:\vspace{-3mm}
\begin{eqnarray*}
    K_i=
    \begin{cases}
      w(1+\epsilon)(1+u)=:k_u, & \text{w.p. (with probability) }\ \delta \\
      w(1+\epsilon)(1+d)=: k_d, & \text{otherwise.}
    \end{cases}
  \end{eqnarray*}
  This is the well-known {\it binomial model (\cite{acemoglu2015systemic,eisenberg2001systemic}), in which the upward moment occurs with probability $\delta$ and downward moment with $(1-\delta).$}
 The agents have to return $y$ (after the interest rate $r_b$) amount to their creditors, however, may not be able to manage the same because of the above economic shocks.  
In case of default,  the agents return the maximum possible;
let $X_i$ be the amount cleared by   the $i$-{th} agent of  group $G_2$.  
 Here we consider a standard   bankruptcy rule, limited liability, and pro-rata basis repayment of the debt contract (see  \cite{acemoglu2015systemic,eisenberg2001systemic}), where the amounts returned are proportional to their liability ratios.
 Thus  node   $j$ of $G_2$  pays back $X_j L_{ji} / y$  towards node $i$,  where  $L_{ji}$   the  liability/the amount borrowed during initial investment phases equals (see the details of previous sub-section and equation (\ref{Eqn_liabOne})): \vspace{-3mm}
\begin{equation}
\label{Eqn_liabilityji}
  \hspace{15mm}  L_{ji}=
    \begin{cases}
      I_{ji} \frac{w}{n p_{ss}}, & \text{if}\ i \in G_1 \\
      I_{ji} \frac{\alpha w(1+\epsilon)}{n p_{ss} (1-\alpha)(1-\epsilon)}, & \text{if}\ i \in G_2.
    \end{cases} 
  \end{equation}Thus the maximum amount cleared by any agent $i \in G_2$, $X_i$,  is given by the following  fixed point equation in terms of the clearing vector 
$\{ X_j \}_{j \in G_2}$ composed    of clearing values of all the agents
  (see \cite{acemoglu2015systemic,eisenberg2001systemic} etc):

\vspace{-3mm}
{\small\begin{equation}
\label{Clearing vector}
 X_i =\min \left  \{  \bigg( K_i+ \sum_{j\in G_2} X_j\frac{L_{ji}} {y} - v \bigg )^+, \    y \right \},
\end{equation}}with the following details: the term $K_i$ is returns of risky investment,  the term     $ \sum_{j\in G_2} X_j\  L_{ji}/y $  equals the claims from the other agents (those borrowed from agent $i$) and $v$ denotes the   taxes/operational costs/deposits/senior liabilities (e.g., as in \cite{acemoglu2015systemic,glasserman2015likely,saha2021random,kavitha2018random}).  In other words, agent $i \in G_2$  will
pay back  the (maximum possible) amount,  $( K_i+   \sum_{j\in G_2} X_j\frac{L_{ji}}{y} - v)^+$,  in case of a default,  and  otherwise, will exactly pay back the  liability amount $y$.

The surplus of any agent  is defined as  the amount left behind, after clearing all the liabilities.  This represents the utility of the agent in the given round. 
The surplus/return of the agent  $i \in G_{2}$ and $i \in G_{1}$ respectively are given by,

\vspace{-4mm}
{\small{\begin{eqnarray}
\label{Retun for G_2}
R^2_i  =  \left (K_i+\sum_{j\in G_2} X_j\frac{L_{ji}}{y} -v-y \right )^+  \hspace{-1mm}\mbox{and } 
\label{Retun for G_1}
R^1_i  =   \left  ( w\epsilon(1+r_s)+\sum_{j\in G_2} X_j\frac{L_{ji}}{y}  -v \right )^+
\hspace{-3mm}. \hspace{3mm}
\end{eqnarray}}}
\vspace{-1mm}
\section{Asymptotic approximation}
\label{sec_aymptotic}
We thus have dynamic graphs whose size increases, with some probability,    in  each round.  In this section, we obtain an appropriate asymptotic analysis of   these graphs, with an aim to derive the pay-off of each group after each round. Towards this, we derive the (approximate) closed form expression of the equation \eqref{Retun for G_2}  which are the returns of the various agents after the settlement of the liabilities. The returns of the agents depend upon  how other agents settle their liabilities to their  creditors.  Thus our first step is to derive the  solution of the clearing vector fixed point equations \eqref{Clearing vector}. Observe that the  clearing vector  $\{X_j\}_{j \in G_2}$ is the solution of the  vector-valued  random fixed point equations \eqref{Clearing vector}  in $n$-dimensional space (where $n$ is the size of the network), defined sample-path wise.

\underline{Clearing vectors using results of \cite{saha2021random,kavitha2018random}:} Our  financial framework can be analysed using the results of   \cite{saha2021random},  
as the details of the model match the assumptions of the paper. Observe that $\alpha(1+\epsilon)/(\alpha+\epsilon) < 1$. By results of  \cite[Section 4]{saha2021random} (like \cite[Theorem 1]{kavitha2018random}), 
 the aggregate claims   converge almost surely  to constant values, as the network size increases to infinity (see subsection \ref{sec_asym_approx} of Appendix for more details): 
 \begin{eqnarray}
 \label{Eqn_aggregate_limitof individualgroup}
 \mbox{(claims of $G_1$)},  &  \displaystyle\sum_{j \in G_2}X_j \frac{L_{ji }}{y}  \to   & \frac{(1-\alpha)(1-\epsilon)}{\alpha+\epsilon} {\bar x}^{ \infty}  \mbox{ a.s.},  \nonumber \\
  \mbox{(claims of $G_2$)},  & \displaystyle \sum_{j \in G_2}X_j \frac{L_{ji }}{y}  \to   & \frac{\alpha (1+\epsilon)}{(\alpha +\epsilon)} {\bar x}^{ \infty} \mbox{ a.s.},
\end{eqnarray}
where  the common expected clearing value ${\bar x}^{ \infty}$ satisfies the following fixed point equation\footnote{
This approximation is valid for $\epsilon \in (0, 1)$. 
In Appendix we also provide the details for $\epsilon\in\{0,1\}$.} in one-dimension (see \cite[Section 4]{saha2021random},  \cite[Theorem 1]{kavitha2018random}):\vspace{-2mm}

\begin{equation}
\label{Eqn_aggregate}
{\bar x}^{ \infty} = E \bigg[\min \left  \{  \bigg( K_i+  \frac{\alpha (1+\epsilon) }{\alpha + \epsilon}  {\bar x}^{ \infty}  - v\bigg)^+,  y  \right \}\bigg].
\end{equation}
Also by the same theorem, the clearing vectors converge  almost surely to (asymptotically independent) random vectors:

\vspace{-4mm}
{\small
\begin{eqnarray}
\label{Eqn_clearingvector_limit}
 X_i \to  \min \left  \{  \bigg( K_i+  \frac{\alpha (1+\epsilon) }{\alpha + \epsilon}  {\bar x}^{ \infty}  - v \bigg)^+,  y \right \}, \mbox{ for  } i \in G_2. 
 \end{eqnarray}}
By virtue of the above results,  the random returns given by  \eqref{Retun for G_2}  converge  almost surely:

\vspace{-4mm}
{\small
\begin{eqnarray}
 R^1_i &\to&  \left  ( w\epsilon(1+r_s)+ \frac{(1-\alpha)(1-\epsilon)}{(\alpha +\epsilon)}  {\bar x}^{ \infty}  -v \right )^+, \mbox{ for } i \in G_1 \nonumber
\\
\label{Return_2}
 R^2_i  & \to &  \left  (K_i+\frac{\alpha (1+\epsilon) }{\alpha + \epsilon}  {\bar x}^{ \infty} -v-y \right )^+,  \mbox{ for} \  i \in G_2.
 \end{eqnarray}}
\underline{Probability of default:} It is defined as  the fraction of agents of $G_2$ that failed to pay back their full liability,  i.e.,  $P_d:= P({ X_i} <y)$. 
For large networks (when the initial network size $n_0 $ itself is sufficiently large), one can use the above approximate expressions and using the same we obtain the  default probabilities and the aggregate clearing vectors in the following (proof is in Appendix):
\begin{lemma}
\label{Lem_Average_clearing}
Consider $w(1+d)(1+\epsilon) \ge  v$. There is a unique solution to  \eqref{Eqn_aggregate} and the asymptotic  average clearing vector and the  default probability of $G_2$  is  given by:

\begin{equation}
\label{Eqn_eps_rnd_star}
(  {\bar x}^{ \infty}, P_d)=
    \begin{cases}
    (y,  \hspace{18mm} 0)                & \text{if }\ \  c_\epsilon  \ge a_1  \\
  \left   (\frac{\delta y +(1-\delta) \underline{w}}{1-(1-\delta)c_\epsilon},       1-\delta \right )     & \text{if} \ \  a_2 \le c_\epsilon < a_1 \\
      \left (\frac{E[W] }{1-c_\epsilon},           \hspace{12mm} 1 \right)                                  & \text{if }  \ \ c_\epsilon < a_2,
    \end{cases}
  \end{equation}where, $c_\epsilon = \frac{\alpha +\alpha \epsilon}{\alpha +\epsilon}$,
$E[W] = \delta \overline{w} + (1-\delta) \underline{w}$, $\underline{w} = k_d -v$,  $\overline{w}= k_u -v$, $a_1 = \frac{y-\underline{w}}{y}$ and $a_2= \frac{y- \overline{w}}{y- (1-\delta)(\overline{w}-\underline{w})}$. 
\eop
\end{lemma}

{\it 
We hence forth, replace \eqref{Retun for G_2} with its approximation \eqref{Return_2}.}
 In \cite{saha2021random} we have performed  Monte Carlo (MC) simulations to demonstrate that the asymptotic approximation well matches the MC estimates with  just few hundred financial entities. The approximation is better for more regular graphs. 
 

\section{Financial Replicator Dynamics}
\label{Sec_finance_replicatordynamics}
In every round of investments, we have a new network, that represents the liability structure of all the agents of that round, formed by their investment choices. 
New agents  may join  the network  and  choose their strategy, as well as the existing agents may  modify their choices in each round.   

The main purpose of  the study of such  replicator dynamics is to derive  asymptotic analysis and answer questions such as: will the dynamics converge, i.e., would the relative fractions of various populations  settle as the number of rounds increase? 
will some of the strategies disappear  eventually? if more than one population type survives what would be the asymptotic fractions? etc.  
 This kind of analysis is common in various other types of networks (e.g., wireless networks (e.g., \cite{MeanWireless}), biological networks (e.g., \cite{ESS})), but is relatively less studied in the context of financial networks (e.g., \cite{LiHonggang}).  We are interested in   asymptotic outcome of these kinds of dynamics (if there exists one) and study the influence of various network parameters on the same. 
 We also study the outcome after finite number of rounds.

\noindent{\bf Dynamics:}
After each round, a random  number of new agents join the network and a random set of agents switch their strategies. 
%
 %
 %
Let $\Nw_t $ be the number of new agents joining the network in round $t$.  Each of them  contact two random  agents, uniformly   from  the ones that participated in the previous round. If both the contacted agents belong to the same group, the new agent adapts the strategy of  that group. Otherwise, {\it the new agent adapts the strategy of the agent that had    higher return in previous round.} The contacted agents   may not reveal  complete information, which can influence the  choices of the new agent.  We assume $\{\Nw_t \}_t$ are i.i.d and that $P({\cal N}_t < {\bar \Nw}) =1$ for some constant ${\bar \Nw}< \infty$.

 Let $(n_1(t),n_2(t), n(t)=n_1(t)+ n_2(t))$  represent respective sizes of  groups $G_1$, $G_2$ and total population.  Let $\epsilon_t  := n_1(t)/n(t) $ represent the fraction  belonging to  $G_1$. The conditional probability and expectation, conditioned on $\epsilon_t$, is denoted by $P_{\epsilon_t}$ and $E_{\epsilon_t}$.
 A new agent joins $G_1$,  if it contacts both the agents from $G_1$  (which happens with conditional probability $\epsilon_t^2$). In case  it contacted one agent from each group (which happens w.p. $2 \epsilon_t (1-\epsilon_t)$), it would join $G_1$ w.p.   $q_{\epsilon_t} := P_{\epsilon_t}(R^1 \ge R^2)$; observe $q_{\epsilon_t}$ is the probability that the return of  agent from $G_1$ is superior. Then the expected number of agents $\xi_t$ that join $G_1$ among $\Nw_t$, conditioned on $\epsilon_t$, is given by:  
$$
E_{\epsilon_t} [\xi_t] = E_{\epsilon_t} \left [E[ \xi_{1}(t) \lvert \Nw_t] \right ] = E[\Nw_t] \left (\epsilon_t^2 +2 \epsilon_t (1-\epsilon_t) q_{\epsilon_t} \right ).
$$

This is the case when the contacted agents reveal correct information. 
{\it Due to incorrect  information, any new agent can make an erroneous decision w.p. $(1-b_n)$.} Thus, 
\begin{eqnarray}
\label{Eqn_expected_numberofswitch}
E_{\epsilon_t} [\xi_t] &=& E[\Nw_t] (\epsilon_t^2 + \epsilon_t (1-\epsilon_t) {\hat q}_{\epsilon_t}),  \mbox{ with, } \nonumber\\
{\hat q}_{\epsilon_t} &:=&  b_n  q_{\epsilon_t} + (1- b_n) (1-{ q}_{\epsilon_t}). \end{eqnarray} 
  



A random i.i.d. number ($\Sw_t $) of  the existing agents may attempt to switch their strategy. Each of them contact another agent  (sampled uniformly) to compare with their experience.  {\it The   strategy of the former is modified to that of the contacted agent's only if the (revealed) return of the  latter is better.}
  
 Any  agent attempting to switch, is from $G_2$ w.p. $1-\epsilon_t$; and it switches to $G_1$ only if the contacted agent is from $G_1$ and if the corresponding return is better. Further due to  incorrect information, there can be an erroneous decision w.p. $(1-b_s)$. In all, any agent among $\Sw_t$ can switch from $G_2$ to $G_1$ with (conditional) probability\footnote{Given that the agent switching is from $G_1$, the probability that the contacted agent is from $G_1$ is actually $\epsilon_t n(t)/(n(t)-1) \approx \epsilon_t$, the approximation improves as $n(t) \uparrow$. }:
 \begin{equation}
 \label{Eqn_transition_switchG_1}
 (1-\epsilon_t) \epsilon_t  {\hat q}^s_{\epsilon_t} \mbox{ where }
  {\hat q}^s_{\epsilon_t}  :=
 \left (b_s q_{\epsilon_t} + (1-b_s) (1- q_{\epsilon_t}) \right ).
 \end{equation}
   Let $\Xi_1(t)$ be the number of agents among $\Sw_t$ that switch from $G_{2}$ to $G_1$, similarly define $\Xi_2(t)$.
 From \eqref{Eqn_transition_switchG_1}, computing as before, $E_{\epsilon_t} [\Xi_1(t)] = \epsilon_t(1-\epsilon_t) \hat{q}_{\epsilon_t}^sE[\Sw_t]$. In similar lines, 
    $$
   E_{\epsilon_t} [\Xi_2(t)] = \epsilon_t(1-\epsilon_t) \left (b_s(1- q_{\epsilon_t}) + (1-b_s)  q_{\epsilon_t} \right ) E[\Sw_t].
   $$
  Thus the expected number of changes in $G_1$ are:
 \begin{eqnarray*}
\label{Eqn_switchtransition}
E_{\epsilon_t}[\Xi_1(t)-\Xi_2(t)  ] &=&
\epsilon_t(1-\epsilon_t)(2 \hat{q}_{\epsilon_t}^s -1) E[\Sw_t], \nonumber\\
& & \hspace{-15mm}= \ \epsilon_t(1-\epsilon_t)(2b_s-1) (2q_{\epsilon_t}-1)E[\Sw_t].
\end{eqnarray*}
   


\noindent{\bf Analysis:}
 The evolution after  round $t$ is summarized by:
\begin{eqnarray}
  \label{eqn_randomdynamic_with_additionandswitch}
  n_1(t+1) &=& n_1(t) + \xi_t + \Xi_1(t) - \Xi_2(t) \mbox{ and, } \nonumber \\
 n(t+1) &= &n(t) + \Nw_t  .
 \end{eqnarray}
 
 
  Let  $\epsilon_0$ and $n_0$ respectively represent the initial fraction and total population  in the system. Define  
$ \psi_{t} := \nicefrac{ n(t)}{(t+n_0)}$ to represent  the rate at which population  evolves, and observe: 
\vspace{-6mm}
 \begin{equation}
  \label{Eqn_psi_update}
      \psi_{t+1}  = \psi_t + \frac{1}{t+n_0+1} \left ( \Nw_t  - \psi_t \right ).
 \end{equation}
 Define  {\small$W_{t+1}  =   \xi_t + \Xi_1(t) - \Xi_2(t).$} Recall   $\epsilon_t =\nicefrac{n_1(t)}{n(t)}$ and so,
 
 \begin{eqnarray*}
\epsilon_{t+1} 
&=&  \frac{n(t) \epsilon_t + W_{t+1} }{n(t)+\Nw_t} \ 
= \ \epsilon_t + \frac{1}{ t+n_0+1} \frac{W_{t+1}  -\Nw_t\epsilon_t}{\psi_{t+1}} .  
  \end{eqnarray*}
  Define 
$\Theta_t = [\epsilon_t, \psi_t]^T$.
One can rewrite the update equations,
\begin{eqnarray}
\label{Eqn_with_additionswitch_updaterule}
\Theta_{t+1} &=&   \Theta_t + \gamma_t Y_{t+1}  \mbox{\normalsize , with, }  \gamma_t\  := \  \frac{1}{t+n_0+1} \ \mbox{\normalsize and }\nonumber \\
Y_{t+1} &:=& \left [ \frac{ W_{t+1} -\Nw_t\epsilon_t}{\psi_{t+1}},  \   \Nw_t - \psi_t \right ]^T\hspace{-2mm}. 
\end{eqnarray}
The conditional expectation of $Y_{t+1}$ with respect to  the $\sigma$-algebra $\mathcal{G}_t := \sigma(\epsilon_s, \psi_{s}, \  s  \le t), $ is given by the following:
\begin{eqnarray}
\label{Eqn_g_beta}
     E[Y_{t+1}\lvert\mathcal{G}_t]  &=& \bar{\bf g} (\epsilon_t, \psi_t) + [e_t,  \ 0],
     \mbox{ with } \bar{\bf g} = [{\bar g}_\epsilon, \bar{g}_\psi], \mbox{\normalsize where,  }
     \nonumber\\
     \bar{g}_\epsilon (\epsilon, \psi)
 &:=& \frac{ \beta\epsilon(1-\epsilon)(2q_{\epsilon}-1)}{\psi} ,  \ \  \bar{g}_\psi (\epsilon, \psi) = E[\Nw]-\psi,  \nonumber \\
\nonumber
e_t &:=& E\bigg[\bigg(\frac{1}{\psi_{t+1}}- \frac{1}{\psi_{t}} \bigg)\bigg(W_{t+1} -\Nw_{t}\epsilon_t \bigg) \Bigg \lvert\mathcal{G}_t \bigg], \mbox{\normalsize and, } \\
\beta &:= &\bigg [(2b_n-1)E[\Nw] + (2b_s-1) E[\Sw]\bigg]. 
\end{eqnarray}
Stochastic approximation methods are well known techniques to study   random trajectory $\{(\psi_{t}, \epsilon_t)\}$ given in \eqref{Eqn_with_additionswitch_updaterule} by approximating it with an ODE formed using the above conditional expectations: 
  \begin{eqnarray}
  \label{Eqn_ODE_with_betap_c}
  \dot{\epsilon} 
  &=& \bar{g}_\epsilon(\epsilon,\psi ), \mbox{ and, } 
  \\
\dot{\psi} &=&  E[\Nw] - \psi \mbox{ for } \ \epsilon\in [0,1], \ \psi > 0 \nonumber.
 \end{eqnarray}
 In Theorem \ref{Thm_RandomC_conv} (given later) we will prove that the above ODE indeed approximates \eqref{Eqn_with_additionswitch_updaterule}, by showing
that
the difference term $e_t$ would converge to 0 (see proof of Theorem \ref{Thm_RandomC_conv}).
  Further,   the right-hand side of the ODE   is not    continuous, because of discrete-valued mapping $\epsilon \mapsto q_\epsilon$, which can take values only among $0, (1-\delta)$ and 1. We nevertheless prove the approximation result and derive further   results using \cite[Theorem 2.2, pp. 131]{kushner2003stochastic}.
Since  ${\bar g}_\epsilon(\cdot)$ is only measurable in $\epsilon$, the results cannot be applied directly. We provide the required justifications while proving Theorem \ref{Thm_RandomC_conv}. 

Prior to that we study the function  $\epsilon \mapsto q_\epsilon$,
  with an aim to analyze
ODE  \eqref{Eqn_g_beta}; this study utilizes the asymptotic approximation derived in
 the previous  section.
The following  results are required for analyzing  the ODE   in   Corollary \ref{Lemma_ODE_Analysis} (provided after Theorem  \ref{Theorem_financenetwork}).

\begin{thm} 
\label{Theorem_financenetwork}
Assume $w(1+d) \ge  v$ for the financial  network. 
\begin{enumerate}[a)]
\item [a)] There exist an  $0\le \bar{\epsilon} \le 1$ such that,    $ q_\epsilon = 1-\delta$  for all $\epsilon < \bar{\epsilon}$ and  $q_\epsilon =1$ for all $\epsilon  \ge  \bar{\epsilon}$.
    \item [b)] We have ${\bar \epsilon} <  1$ if and only if (with $w_{dv} := w(1+d)-v$)
\begin{eqnarray}
 \label{Eqn_cond_bareps}
 \frac{  w (r_b-d) - w_{dv} } {w( 1+2\alpha\delta- \alpha)} &>& \frac{(2u-r_s-r_b)}{2\alpha(1-\delta)} \mbox{ or } \\
 \frac{ w (r_b-d) - w_{dv} }{w(1+2\alpha\delta- \alpha)}   &>&  \frac{ 2   (u-d) }{1+\alpha}
 \mbox{ and }  \frac{u-r_b}{u-d} < \alpha   (1-\delta). \nonumber
 \end{eqnarray}
 \item [c)] The ${\bar \epsilon}$ can be derived as zero of the   equation:
  \begin{eqnarray*}
 \hspace{-6mm}
 \epsilon^2 + m_1 \epsilon + m_2 = 0,  \mbox{ with, } m_2=  \frac{ w\alpha \delta(u-r_b)+(2\alpha-1)(1-\delta)w_{dv}}{m_3}\\
  \hspace{-2mm}m_1 = \frac{ w(u-r_b)(1-\alpha+ \alpha\delta))+ w\alpha \delta(u-r_s)  }{m_3} \hspace{25mm} \\
  +\frac{(1-\delta)\big(w_{dv}-w(r_b-d)(2\alpha-1)\big) }{ m_3}, \hspace{20mm} 
 \end{eqnarray*}
 where {\small$\ m_3 = w(u-r_s)(1-\alpha(1-\delta)) -w(1-\delta)(r_b-d).$} \eop
\end{enumerate}
\end{thm}
Clearly, from ODE \eqref{Eqn_ODE_with_betap_c},  the solution of $\psi$ component is {\small$\psi (t)= E[\Nw]+(\psi_0-E[\Nw]) e^{-t}$}, with $\psi(0) = 1$. Further using Theorem \ref{Theorem_financenetwork}, the $\epsilon$-component of ODE  \eqref{Eqn_ODE_with_betap_c} simplifies to:
\begin{eqnarray}
 \label{Eqn_ODE_epsilon}
  \dot{\epsilon}  =    \frac{   \kappa \epsilon (1-\epsilon)  } { E[\Nw]+(\psi_0-E[\Nw]) e^{-t}  }    \     \mbox{ with, } 
  \ 
  \kappa  \ = \ 
    \begin{cases}
  \beta(1-2\delta)
        & \text{ if }\  \epsilon_t  <  {\bar \epsilon}
        \\ \beta  & \text{ if }\   \epsilon_t \ge {\bar \epsilon}.
    \end{cases}
    \end{eqnarray}
This ODE has 
 unique solution for any initial condition~$\epsilon_0$: 
 \begin{eqnarray}
 \label{Eqn_ODE_solution}
  \epsilon(t) =   \frac{  \epsilon_0 \big ( E[\Nw] e^t + 1-E[\Nw]  \big )^{\kappa/E[\Nw]}  } {1-\epsilon_0 + \epsilon_0 \big ( E[\Nw] e^t + 1- E[\Nw]  \big )^{\kappa/E[\Nw]}  } . 
\end{eqnarray}
    
The above immediately implies  the following result:

\begin{cor}
\label{Lemma_ODE_Analysis}
{\bf [ODE Analysis]}
Assume initial condition $\epsilon(0)  \in (0, 1)$. Then with ${\bar \epsilon}$ as in Theorem \ref{Theorem_financenetwork},

\begin{enumerate}
\item [a)] When $\beta>0$ and $\delta > 1/2$, we have  the following   convergence of the unique ODE solution (as $t \to \infty$):

\vspace{-4mm}
{\small 
\begin{eqnarray}
\label{Eqn_DOA}
\hspace{-4mm}
 \epsilon(t) \to 0  \ \mbox{ if } \ \epsilon(0) < \bar{\epsilon} \mbox{ and } \epsilon(t) \to 1  \mbox{ if } \epsilon(0) \ge  \bar{\epsilon} \ \  (\mbox{actually } \epsilon(0) >  0, \mbox{ when, }\bar{\epsilon}=0)  . \hspace{4mm}
\end{eqnarray}}
\item [b)]  When $\beta <  0$, $\delta > 1/2$ and   $\bar{\epsilon} \in \{ 0,1 \}$, then as $t \to \infty$
\begin{eqnarray*}
  \epsilon(t)  \to 1 \  \mbox{ if } \ \bar{\epsilon} = 1 \  \mbox{ and } \ \epsilon(t)  \to 0 \ \mbox{ if }     \bar{\epsilon} = 0. 
\end{eqnarray*}

\item [c)] When $\delta < 1/2$, $\epsilon(t) \to 0$ if $\beta < 0$, else $\epsilon(t) \to 1$ as $t \to \infty$. \eop
\end{enumerate}
\end{cor}
From \eqref{eqn_randomdynamic_with_additionandswitch}-\eqref{Eqn_with_additionswitch_updaterule},  
$
   \epsilon_k \in [0, 1]$    for all  $k$   and all sample paths. The $\psi $ component can be bounded a.s.   by law of large numbers.
By Corollary \ref{Lemma_ODE_Analysis} the combined domain of attraction of the  asymptotically stable attractors among $\{0, 1\}$ is $[0,1]$   and so  the dynamics  trivially visits the combined domain infinitely often, under the assumptions of the Corollary.
 \emph{We now have the main result}  (proof is in Appendix):
\begin{thm}
\label{Thm_RandomC_conv}
Assume $w(1+d) \ge v.$
The sequence $\{\epsilon_k\}_k$ generated by    \eqref{eqn_randomdynamic_with_additionandswitch}-\eqref{Eqn_with_additionswitch_updaterule} converges almost surely     
to  limit points $\{\epsilon^{*}\}$ of the ODE \eqref{Eqn_ODE_with_betap_c}, obtained in Corollary \ref{Lemma_ODE_Analysis}, under the respective conditions provided in the Corollary. \eop  
\end{thm}
\noindent {\bf Remarks} We will show below that   the  limit points  of Theorem \ref{Thm_RandomC_conv} are ESS and  thus the  dynamics   settle (almost surely, a.s.) to a pure  ESS ($\epsilon^* \in \{ 0, 1\}$), further when switching is predominant.
   
 The   settling    point (ESS), depends upon  network parameters and the initial proportion of agents.
%
 There is a possibility of 
 the dynamics converging to $1$, i.e., 
 $\epsilon_k \to 1$, only when the conditions of Theorem \ref{Theorem_financenetwork}.(b) are satisfied. 
 In all other cases, the system converges to pure ESS with all risky agents.

 
Under the conditions of  Theorem \ref{Theorem_financenetwork}.(a), when $\epsilon >{\bar \epsilon}$  the financial network could have entered systemic risk regime (scenario in which majority of the agents   have defaulted), if further $\epsilon > {\bar \epsilon}_2$ by Lemma \ref{Lemma_mono_PD} provided in Appendix. But the  {\it agents averted such a catastrophic event by all of them choosing the  less risky strategy whenever $\beta >0$; by Corollary \ref{Lemma_ODE_Analysis}, 
$\epsilon(t) \to 1$ for such initial conditions, which reflects the same for random trajectory $\{\epsilon_k\}$. In fact, we noticed that the system always avoids systemic risk regime 
even in 
  test cases with $\beta < 0$,   in  Section \ref{Section_numerical_observations} on numerical study; furthermore, this was true even when $w(1+d) < v$, but we need the obvious condition $v < w(1+u)$. 
  
  Our theoretical study is under the assumption that agents are always able to repay  the  taxes/senior debt (even with downward return,    $w(1+d) > v$). One can easily extend this study for the case when  $w(1+d) <  v < w(1+u)$, we conjecture that \eqref{Eqn_ODE_solution} again represents the ODE solution.  We have few such examples in Figure \ref{fig:perfect_imperfect} where the ODE solution well matches the random trajectory.

}

\noindent{\bf Mixed Limit:}
Consider the case with $w(1+d) \ge  v$ , $\beta < 0$ and     $0 < \bar{\epsilon} < 1$. From \eqref{Eqn_ODE_solution}, it is clear that ODE solution trajectory is 
increasing when below ${\bar \epsilon}$ and decreasing when above ${\bar \epsilon}$.
Thus it   wanders around ${\bar \epsilon}$, as $t \to \infty$. {\it We conjecture that   the finance dynamics converges to this intermediate limit point.}  We indeed illustrate the same using numerical examples in Section \ref{Section_numerical_observations} (see Figure \ref{fig:perfect_imperfect} and Table \ref{Table_MC _basedwithrandom addition_random}).


\noindent{\bf Finite Round Approximation:} 
Using similar arguments as 
in Theorem \ref{Thm_RandomC_conv}, one can show that the sequence  $\{\epsilon_k\}_{k \ge {\bar k}}$  can be approximated by ODE solution uniformly over any finite time  horizon i.e., for  any $T >0$ (e.g., as in \cite[Theorem 1]{singh2021evolutionary}), as ${\bar k} \to \infty$.  To be more precise,  for any $k \ge 1$, we have that:
  \begin{equation}
  \label{Eqn_ode_approx}
  \epsilon_{l+k}  \approx   \frac{  \epsilon_0\big ( E[\Nw] e^{t_{k,l}} +1- E[\Nw]  \big )^{\kappa/E[\Nw]}  } {1-\epsilon_0 + \epsilon_0\big ( E[\Nw] e^{t_{k,l}} + 1- E[\Nw] \big )^{\kappa/E[\Nw]}  },  \  \mbox{\normalsize with, } 
 \end{equation}
  $t_{k,l}  =  \sum_{j=l+1}^{l+k} \frac{1}{j+n_0+l}
  $, when $(l+n_0)$   is sufficiently large;
thus one can   estimate the approximate fraction of entities that adapt the  less risky strategy after $l+k$ number of rounds. In Section \ref{Section_numerical_observations}, we  observe that the approximation is good for majority of cases, even for  $l \approx 1$ and $n_0$ in few hundreds.\\
\textbf{Average Dynamics:} 
 In our previous work \cite{Saha}, we consider the analysis with a large number of samples/observations; we would like to summarize relevant  results of \cite{Saha}, with an aim to  compare them with  the dynamics of the current paper. 
 
Any new agent  at first  contacts  two random (uniformly sampled) agents    of the previous round.   If both the contacted agents belong to the same group, the new agent adapts the strategy of that group. When it contacts agents from both the groups \emph{it investigates more before making a choice}; the new  agent observes significant  portion of the network, in that,  it  obtains a good estimate of  the average utility of agents belonging to both the groups. It adapts the strategy of the group with maximum (estimated) average utility.
 %
 %
 Say it observes the average of each group with an error that is normally distributed with mean equal to  the expected  return  $\phi_i (\epsilon)= E[R_i^{k}] $  with $k=1,2$  and variance proportional to the size of the group, i.e., 
it observes (here ${\cal U}(0, \sigma^2)$ is a zero mean Gaussian random variable  with variance $\sigma^2$)

$
{\hat \phi}_i (\epsilon) =\phi_i (\epsilon) + {\cal U}_i  \mbox{ with } {\cal U}_1  \sim  {\cal U} \left ( 0,  \frac{1}{{\bar c} \epsilon } \right  ) \mbox{ and }  {\cal U}_2  \sim  {\cal U} \left ( 0,  \frac{1}{  {\bar c} (1- \epsilon )  } \right ), 
$
for some ${\bar c}$ large. The dynamics converges to the following limit points (more details  of the dynamics and proof can be found in \cite{Saha}):
 \begin{lemma}
 \label{Lem_avg_dyn}
 Define ${\bar r}_r  := u \delta + d (1-\delta)$. 
 Assume $\epsilon_0 \in (0, 1).$
 There exists a ${\bar \delta} < 1$   (depends upon the instance of the problem) such that the following  are valid for all $\delta \ge {\bar \delta}$: 
 \begin{enumerate}
 \item [a)]If  ${\bar r}_r  >  r_b > r_s$,  then  $\epsilon_k \to  0 $ almost surely (as $k\to \infty$).  
 \item[ b)] If  $\phi_1 (\epsilon) > \phi_2 (\epsilon) $ for all $\epsilon$,   then $\epsilon_k \to 1 $ almost surely. 
 \item [ c)] When $ r_b > {\bar r}_r  >  r_s$,   and case (b) is negated    there exists a unique zero $\epsilon^*$ of the equation $\phi_1(\epsilon) - \phi_2(\epsilon) = 0$  and  
   $$  
   \hspace{-4mm} \epsilon_k \to  \epsilon^{*}  \mbox{  almost surely;   further  for $\delta \to 1$,  }  \epsilon^{*}  \to   \frac {r_b - {\bar r}_r} { {\bar r}_r  - r_s }. \hspace{02mm} \mbox{\eop} 
   $$   
 \end{enumerate}
\end{lemma}
\textbf{Remarks:} 
 From Theorem \ref{Thm_RandomC_conv} and Corollary \ref{Lemma_ODE_Analysis}, under the corresponding hypothesis, the random dynamics always converge to either  `risky' or `less risky' agents, i.e., pure strategies (with $\beta >0$). On the other hand, the average
dynamics, as in Lemma \ref{Lem_avg_dyn}, either converges to pure or mixed strategies further based on the parameters. 
Thus there is a big difference in the settling behaviour depending upon the number of (perfect) observations; 
{\it when agents observe sparsely, the network eventually
settles to one of the two strategies, and if they observe more samples there is a possibility of the emergence of mixed limit.}  
On the other hand, even with sparse observations one can have mixed limit (${\bar \epsilon}$ of Theorems \ref{Theorem_financenetwork}-\ref{Thm_RandomC_conv}) due to imperfect information (see also mixed ESS related discussions in Section \ref{sec_ess}).

\ignore{
From Theorem \ref{Thm_RandomC_conv} and Corollary \ref{Lemma_ODE_Analysis} the random dynamics always converge to either  `risky' or `{\color{red} less risky}' agents, i.e., pure strategies (with $\beta >0$). On the other hand, the average
dynamics, as in Lemma \ref{Lem_avg_dyn}, either converges to pure or mixed strategies further based on the parameters. 
Thus there is a big difference in the settling behaviour depending upon the number of (perfect) observations; 
{\it when agents observe sparsely, the network eventually
settles to one of the two strategies, and if they observe more samples there is a possibility of the emergence of mixed limit.}  
On the other hand, even with sparse observations one can have mixed limit (${\bar \epsilon}$) due to imperfect information. 
}

 \section{Limit points and ESS}
 \label{sec_ess}
We will now show  that the limit points   of the ODE \eqref{Eqn_ODE_with_betap_c}, which are also the (a.s.) limits of the replicator dynamics (by Theorem \ref{Thm_RandomC_conv})  are indeed  evolutionary stable, {\it under predominant switching, i.e., when $sign (\beta) = sign  (2b_s-1) $.} We will also show that the mixed limits  are evolutionary stable.  

The utility (a.k.a. fitness) of any agent is $+1$ if it encounters opposite group agent and if its return is better; it is $-1$, if the opponent has better return. Thus the utility of agent using  less risky strategy, conditioned on $\epsilon_t=\epsilon$:
\begin{eqnarray}
\label{Eqn_agent_utlity1}
 u_1 ( \epsilon) = (1-\epsilon) 
\left (  (b_s q_\epsilon + (1-b_s) q^c_\epsilon)  -  ( b_s q^c_\epsilon + (1-b_s) q_\epsilon)   \right )  \hspace{-5mm}
\nonumber
\\
= (1-\epsilon) (2q_\epsilon - 1) (2b_s - 1) \mbox{, with, } q^c_\epsilon := 1-  q_\epsilon. \hspace{1mm}
\end{eqnarray}
Similarly that of the agent using risky-investments equals:
\begin{equation}
\label{Eqn_agent_utlity2}
u_2 (\epsilon) = - \epsilon (2q_\epsilon - 1) (2b_s - 1).
\end{equation}
\subsection*{Mixed ESS:}
Intuitively  a strategy   (chosen by majority) is called    mixed ESS if a small fraction of mutants using some other  strategy are eventually wiped out. 
We provide the precise definition directly in our finance network context  and show   that all our limits    \eqref{Eqn_ODE_with_betap_c} are indeed mixed ESS. 
 
  For  such definitions one needs to consider mixed strategies; \emph{$\epsilon$ is a mixed strategy, if an agent invests in  less risky  or risky assets respectively with probabilities $\epsilon$ and $1-\epsilon$.} Each agent adapting a mixed strategy $\epsilon$ is equivalent to $\epsilon$ fraction of them investing in  less risky assets (see e.g., \cite{easley2010networks}); for large networks this equivalence is simply due to law of large numbers.
Consider a mixture of  population, where  $x$ (a small) fraction of mutants use $\epsilon$ while the remaining use ${\bar \epsilon}$ (intermediate limit in our case).  With such a  population mix, the fraction of agents investing in  less risky assets is given by,
$
\epsilon_{x}:= x\epsilon +(1-x) \bar{\epsilon}.
$ Let $u(\epsilon, \epsilon_x)$, $u({\bar \epsilon}, \epsilon_x)$ respectively represent the utilities of the mutants and the other agents (non-mutants),  against population mix, $\epsilon_x$. 
\begin{definition}
A  strategy   ${\bar \epsilon}$ is mixed ESS if for any   $\epsilon \ne {\bar \epsilon}$,  there exist ${\bar x} > 0$ such that   
$u(\epsilon, \epsilon_x) < u({\bar \epsilon}, \epsilon_x)$ for all  $x \le {\bar x}$.
\end{definition}
From \eqref{Eqn_agent_utlity1} -\eqref{Eqn_agent_utlity2} the utility of the mutant, i.e., the agent using mixed strategy $\epsilon$,  is given by:

\vspace{-4mm}
{\small 
\begin{eqnarray*}
 u(\epsilon, \epsilon_x) &=& \epsilon u_1 (\epsilon_x) + (1-\epsilon) u_2 (\epsilon_x) \\
 && \hspace{-16mm}= \ \epsilon(1-\epsilon_x)(2q_{\epsilon_x}-1) (2b_s-1)   -(1-\epsilon)\epsilon_x(2q_{\epsilon_x}-1) (2b_s-1).
  \end{eqnarray*}}
One can compute $u({\bar \epsilon}, \epsilon_x)$ similarly, and 
   the difference  in the two utilities:
 \begin{eqnarray}
 \label{Eqn_utility_difference}
  u(\epsilon, \epsilon_x) - u(\bar{\epsilon}, \epsilon_x)=  (\epsilon-\bar{\epsilon})(2q_{\epsilon_x}-1)(2b_s-1).
 \end{eqnarray}
For predominant switching, the mixed limit  is possible  only when $\beta < 0$, i.e., if $2b_s -1 < 0$. Further when $\epsilon < {\bar \epsilon}$, we have $\epsilon_x< {\bar \epsilon}$  for any  $x\in (0, 1)$; thus  from Lemma \ref{Lemma_threshold of q_eps},  $q_{\epsilon_x} = 1-\delta$ and so RHS  of 
  \eqref{Eqn_utility_difference}  is negative. When $\epsilon > {\bar \epsilon}$, again from  Lemma \ref{Lemma_threshold of q_eps},  $q_{\epsilon_x} = 1$ and  the RHS is  negative. Thus ${\bar \epsilon}$ is mixed ESS, with any ${\bar x} $.

  Next   consider   pure limits $\epsilon^{*} \in \{0,1\}$. We will indeed show that there are again Mixed ESS (stable against mutants using mixed strategies), which also implies stability against pure strategies. We again consider  predominant switching. 
  
  From Theorem \ref{Theorem_financenetwork}.(a), we have  $\epsilon^* = 0$ with $\beta >0$,  (i.e., $(2b_s - 1) >0$) and $\delta > 1/2$, only   when $ {\bar \epsilon} > 0$. 
  Thus, 
  for any mutant using (mixed) strategy $\epsilon > 0$,
  by  Lemma  \ref{Lemma_threshold of q_eps}, one can choose small enough ${\bar x}$ such that $q_{\epsilon x} = 1- \delta$ for all $x \le {\bar x}$; basically we require $\epsilon{\bar x} < {\bar \epsilon}$. Then as in \eqref{Eqn_utility_difference}, for all such $x$, 
  
  \vspace{-4mm}
  {\small$$u(\epsilon, \epsilon x) - u(\epsilon^*, \epsilon x) =u(\epsilon, \epsilon x) - u(0, \epsilon x) = \epsilon (1-2\delta) (2 b_s - 1)  < 0.$$} 
  
  Similarly when $\epsilon^* = 1$, the only possible mutant strategies are $\epsilon < 1$,  again  for small enough $x$ we have  
  $q_{\epsilon x + (1-x)} = 1$; the rest of the arguments are similar. 
 One can prove  other pure limits (e.g.,  with $\beta < 0$) are mixed ESS in a similar way.  
 
 When the dynamics are dominated by addition of new agents, then limit points need not be evolutionary stable. That is anticipated as the dynamics are not due to the existing agents; further practically one might have very few additions in comparison to the changes in strategies of the agents. 
 
{\bf Evolutionary stable strategy for average dynamics:}
 The ODE for average dynamics is given by  (see \cite{Saha})
 $$
\dot{\epsilon}= 
  \epsilon (1-\epsilon) \left ( 2 g (\epsilon)  - 1 \right ) \mbox{ where } g(\epsilon) :=  \int_{-\infty}^{  \left ( \phi_1 (\epsilon ) -  \phi_2 (\epsilon) \right )  \sqrt{{\bar c} \epsilon (1-\epsilon) }   }  e^{- x^2 / 2}  \frac{dx }{\sqrt{2 \pi}}.
 $$
 In \cite{Saha} we found the limit points of the ODE and omitted the details regarding them being evolutionary stable. These limit points are  also  summarized in Lemma \ref{Lem_avg_dyn}. 
 We will show below that these limit points are  indeed  evolutionary stable as in the above paragraphs, but now with the (approximate) expected return  \eqref{Return_2} being the utility of any agent.

 \noindent  \textbf{Mixed ESS with  $\epsilon^{*}=  \frac {  r_b - {\bar r}_r  } {  {\bar r}_r  - r_s }$:}  We again consider a mixture of  population, where  $x$ (a small) fraction of mutants use $\epsilon$ while the remaining use $ \epsilon^{*}$.
 From \eqref{Return_2} the utility of the mutant, i.e., the agent using mixed strategy $\epsilon$,  is given by:
 \begin{eqnarray*}
 u(\epsilon, \epsilon_x) &=& \epsilon \phi_1 (\epsilon_x) + (1-\epsilon) \phi_2 (\epsilon_x) \mbox{ where, }
 \epsilon_x= x\epsilon +(1-x)\epsilon^{*}. 
 \end{eqnarray*}
 One can compute $u(\epsilon^{*}, \epsilon_x)$ similarly, and the difference  in the two utilities:
 \begin{eqnarray}
 \label{Eqn_average_dynamics_utilitydifference}
  u(\epsilon^{*}, \epsilon_x) - u(\epsilon, \epsilon_x)= (\epsilon^{*}-\epsilon)(\phi_1(\epsilon_x) -\phi_2(\epsilon_x)).
 \end{eqnarray}
  When $\epsilon < \epsilon^{*}$, we have  $\epsilon_x< \epsilon^{*}$ for any $x \in (0,1)$; thus from   the proof of \cite[Corollary 1, pp. 227]{Saha}, $\phi_1(\epsilon_x)  > \phi_2(\epsilon_x)$ (when $\delta \ge {\bar \delta}$), and hence the RHS of \eqref{Eqn_average_dynamics_utilitydifference} is positive.
  Similarly, the RHS is positive  even when $\epsilon >  \epsilon^{*}$. Hence $\epsilon^{*} = \frac {  r_b - {\bar r}_r  } {  {\bar r}_r  - r_s }$ is a mixed ESS under the conditions of Lemma \ref{Lem_avg_dyn}(c).

\noindent  \textbf{ Pure ESS with  $\epsilon^{*}= 0$:} For any mutant using (mixed) strategy $\epsilon > 0$,
    by  Lemma \ref{Lem_avg_dyn}(a) and \eqref{Eqn_average_dynamics_utilitydifference}, for all such $x$, 
 \begin{eqnarray}
 \label{Eqn_utilitityofG1age}
 u(\epsilon^{*}, \epsilon_x) - u(\epsilon, \epsilon_x) &=& \epsilon (\phi_2(\epsilon_x) - \phi_1(\epsilon_x)) > 0.
\end{eqnarray}
  Therefore $\epsilon^{*}= 0$ is an ESS under the conditions of Lemma. \ref{Lem_avg_dyn}(a). 
  
 \noindent  \textbf{Pure ESS with 
 $\epsilon^{*}= 1$:} Similarly when $\epsilon^* = 1$, the only possible mutant strategies are $\epsilon < 1$, again by  Lemma \ref{Lem_avg_dyn}.(b) and \eqref{Eqn_average_dynamics_utilitydifference}, for all such $x$,
  \begin{eqnarray}
  \label{Eqn_utilitityofG2age}
  u(\epsilon^*, \epsilon x) - u(\epsilon, \epsilon x)  = (1-\epsilon) (\phi_1(\epsilon_x) - \phi_2(\epsilon_x)) > 0. 
  \end{eqnarray}
  Therefore $\epsilon^{*}= 1$ is an ESS  under the conditions of Lemma \ref{Lem_avg_dyn}.(b).

   Thus the replicator dynamics either settles to a pure strategy ESS or mixed ESS (see Lemma \ref{Lem_avg_dyn}.(c)),  depending upon the parameters of the network;
  after a large number of rounds, either the fraction of agents following one of the strategies converges to one or zero  or  the system reaches a mixed ESS which balances the expected returns of the two groups.

{\bf Stability against multiple mutations:} 
Towards the end,   we would like to comment on the stability against mutations (e.g., \cite{ghatak2012evolutionary}). In this case, at a time,  multiple mutant strategies can prevail and one needs to show that the strategy under consideration is stable. Basically if say $\epsilon_1, \epsilon_2, \cdots, \epsilon_k$  are the mutant strategies that exist in proportions $x_1, x_2, \cdots, x_k$  and $\epsilon^*$ is the candidate for ESS.  Then $\epsilon^*$ is said to be stable against multiple mutants  (e.g., \cite{ghatak2012evolutionary}), if there exists a threshold ${\bar x}$ such that:  when $\sum_i x_i \le {\bar x}$,   
$$
u( \epsilon^*, \epsilon_x ) > u (\epsilon_i, \epsilon_x) \mbox{ for all }  i,
$$ where $\epsilon_x = \sum_i \epsilon_i x_i +  (1- \sum_i x_i) \epsilon^*$ is the mixed strategy after mutants  (as defined above). As shown in \cite{ghatak2012evolutionary}, only pure strategies can be stable against multiple mutations. Thus our mixed ESS ${\bar \epsilon}$ would not be stable against multiple mutations. However one can easily verify that the pure strategies are indeed stable against multiple mutations also (see \eqref{Eqn_utilitityofG1age}-\eqref{Eqn_utilitityofG2age}). 

\section{Defaulters may stop investing}
\label{Section_defaultersleave}
We now consider the scenario where the agents can stop investing. To be more specific, the defaulted agents (or some of them) would stop investing, after each round. This represents a more realistic scenario and our aim is to compare the asymptotic outcome of such a system with that of     the previous sections.   
 
 \noindent{\bf Dynamics:} As before, a random number of new agents join the network, and a random subset of agents switch their strategies; these details are the same as in the previous sections.  Additionally, a random number of defaulted agents  leave the network.   Let $\Dw_k$ be the number of the defaulted  agents that stop investing after round $k$.

\noindent{\bf Analysis:}
The evolution after  round $k$ is  given by (notations   as in  \eqref{eqn_randomdynamic_with_additionandswitch}-\eqref{Eqn_g_beta}):

 \vspace{-4mm}
 {\small \begin{eqnarray}
\label{Eqn_randomdynamic_with_additionandswitchdefault}
  n_1(k+1) = n_1(k) + \xi_k + \Xi_1(k) - \Xi_2(k), \mbox{ and, }   
 n(k+1) \ = \ n(k) + \Nw_k - \Dw_k. \hspace{4mm}
 \end{eqnarray}}
Recall there would be no defaults in the group $G_1$.

With the remaining details exactly as in Section \ref{Sec_finance_replicatordynamics}, given by equations \eqref{eqn_randomdynamic_with_additionandswitch}-\eqref{Eqn_g_beta}, the fraction  $\psi_{k} = \nicefrac{ n(k)}{(k+n_0)}$  evolves according to the following:
\begin{equation}
\label{Eqn_psi_updatewithdefault}
 \psi_{k+1}  = \psi_k + \frac{1}{k+n_0+1} \left ( \Nw_k -\Dw_k  - \psi_k \right),
\end{equation}
 and $\Theta_k = [\epsilon_k, \psi_k]^T$  with $\epsilon_k := \nicefrac{n_1(k)}{n(k)}$ evolves as below:
 
 \vspace{-3mm}
 {\small
\begin{eqnarray}
\label{Eqn_with_additionswitch_updaterulewithdefault}
\Theta_{k+1} =   \Theta_k + \gamma_k Y_{k+1}  \mbox{, with, }  
Y_{k+1} := \left [ \frac{ W_{k+1} -\Nw_k\epsilon_k+ \Dw_k \epsilon_k}{\psi_{k+1}},  \  \ \Nw_k - \Dw_k- \psi_k \right ]^T. \nonumber  
\end{eqnarray}}
The conditional expectation of $Y_{k+1}$ with respect to   $\sigma$-algebra $\mathcal{G}_k := \sigma(\epsilon_s, \psi_{s}, \  s \le k) $ and its error $e_k$ have similar  structure as before:
\begin{eqnarray}
\label{Eqn_key_dynamicsdefault}
     E[Y_{k+1}\lvert\mathcal{G}_k]  &=& \bar{\bf g}^D (\epsilon_k, \psi_k) + [e_k, 0]^T, \mbox{ with, }   \\
     \bar{\bf g}^D (\epsilon_k, \psi_k) &:=&  \left [ \bar{g}^D_\epsilon (\epsilon_k, \psi_k)  ,  \   \bar{g}^D_\psi (\epsilon_k, \psi_k)    \   \right ]^T ,   \nonumber
     \\
       e_k &:=& E\bigg[\bigg(\frac{1}{\psi_{k+1}}- \frac{1}{\psi_{k}} \bigg)\bigg(W_{k+1} -(\Nw_{k} -\Dw_{k})\epsilon_k \bigg) \Bigg \lvert\mathcal{G}_k \bigg],  \nonumber \\
        \bar{g}^D_\epsilon (\epsilon, \psi)
 &:=& \frac{ \beta \epsilon(1-\epsilon)(2q_{\epsilon}-1) +  \epsilon E_\epsilon [\cal D] }{\psi}  \stackrel{a}{=} \frac{ \kappa \epsilon(1-\epsilon)  +  \epsilon E_\epsilon [\cal D] }{\psi}, \mbox{ and, }\nonumber \\
   \bar{g}^D_\psi (\epsilon, \psi) &:=& E[\Nw]- E_\epsilon [{\cal D}] -\psi, 
   \mbox{ with } \kappa = 
   \beta 1_{\epsilon \ge {\bar \epsilon}} + \beta (1- 2\delta)  1_{\epsilon < {\bar \epsilon}},
   \nonumber 
\end{eqnarray}  
where  the remaining terms are  as in \eqref{Eqn_g_beta} and equality $a$ follows by Lemma \ref{Lemma_threshold of q_eps}. 
%

We consider that at maximum, an i.i.d. number of defaulters stop investing after each round.  Recall that $n_2(k)$ is the number of $G_2$ agents, and $P_d=P_d(\epsilon_k)$ is the probability that a typical $G_2$ agent defaults. 
 Thus one can  model  $\Dw_k \sim \min \{{\cal L}_k,  Bin(n_2(k), P_d )\}$  as the number of defaulters leaving the system\footnote{We basically assume each agent among $n_2(k)$  number of the $G_2$ agents default (asymptotically) independently of each other and this is valid by \cite[Section 4]{saha2021random},  \cite[Theorem 1]{kavitha2018random}, as the clearing vectors are asymptotically independent.} conditioned on $\epsilon_k$; here $\{ {\cal L}_k\}$ is an i.i.d sequence and assume $P({\cal L} \le {\bar {\cal L}}) = 1$, for some $\ {\bar {\cal L}} < \infty.$ 
For this case study from equation \eqref{Eqn_key_dynamicsdefault},  when the number of rounds $k$ is sufficiently  large,   the random trajectories can be  approximated by  the ODE as below:\vspace{-4mm}
\begin{eqnarray}
\label{Eqn_odewithdefaultiidbounded}
 \dot{\epsilon} 
  &=& \bar{g}^D_\epsilon (\epsilon, \psi), \mbox{ and, }  \nonumber \ 
\dot{\psi} \ = \ 
E[\Nw] -   E_\epsilon[\mathcal{D}]   - \psi \mbox{, where, }  \\
E_\epsilon[\mathcal{D}] 
&=& 1_{\epsilon < 1} 1_{ P_d(\epsilon) > 0} E[\mathcal{L}]  .
\end{eqnarray}
The last equality follows by boundedness of $\{{\cal L}_k\}$ sequence;  it may appear that $E_\epsilon[\mathcal{D}_k] $ depends upon $k$, however by boundedness of $\{{\cal L}_k\}$,  we have $E_\epsilon[\mathcal{D}] 
= 1_{\epsilon < 1} 1_{ P_d(\epsilon) > 0} E[\mathcal{L}] $ at limit ($k\to \infty$). 
By Lemma \ref{Lemma_mono_PD} of Appendix, 
the ODE equation \eqref{Eqn_odewithdefaultiidbounded}    simplifies  to the following for $\epsilon < 1$ (see equation \eqref{Eqn_ODE_epsilon}):
\begin{eqnarray}
\label{Equation_epsilon_simplified_equation}
\dot{\epsilon}&=&
 \begin{cases}
 \frac{ \kappa\epsilon(1-\epsilon)}{\psi} 
        & \text{ if }\  \epsilon < \bar{\epsilon_1}, \mbox{ with } {\bar \epsilon_1}=\frac{w(1+d)-v}{w(r_b-d)}  \nonumber \\  
        \frac{\kappa  \epsilon(1-\epsilon) }{\psi}  +\frac{\epsilon E[\cal{L}]}{\psi}
        & \text{ else.} \end{cases}
\\
\dot{\psi}&=&
 \begin{cases}
 E[\Nw] -\psi
        & \text{ if }\  \epsilon < \bar{\epsilon_1}, \\  
       E[\Nw] -\psi - E[\mathcal{L}]
        & \text{ else. } 
\end{cases}
\end{eqnarray}

We begin with the ODE approximation result, which requires the following assumption. After the theorem, we  show that the assumption is satisfied. 
\begin{enumerate}
    \item[{\bf A.}] Let the set ${\cal A}$ be locally asymptotically stable\footnote{See Definition \ref{Def_Asymptotically stable} of Appendix, for  the definition of the asymptotically stable attractor. } in the sense of Lyapunov for the ODE \eqref{Equation_epsilon_simplified_equation}.  Assume that $\{\up_k\}$, with $\up_k := (\epsilon_k, \psi_k)$ visits a compact set, $S_A$, in the domain of attraction (DoA), ${\cal D}_A$, of ${\cal A}$ infinitely often (i.o.) with probability  $\varrho > 0$.
\end{enumerate}
\begin{thm}
\label{Thm_RandomC_convwithdefault}{\bf [ODE     Approximation]}
Assume as in Theorem \ref{Thm_RandomC_conv}, and  further assume  $E[\Nw]> E[\mathcal{L}]$. 
\begin{enumerate}
\item[i)]   For every $T>0$, almost surely there exists a sub-sequence $(k_m)$ such that: 
        $$
        \hspace{-6mm}
            \sup_{k: t_k \in [t_{k_m}, t_{k_m} + T]} d(\up_k, \up( t_k - t_{k_m})) \to 0,  \mbox{ as } m \to \infty, \mbox{ where, } t_k:=\sum_{i=1}^{k}\gamma_i \mbox{, and,}
            $$
        $\up (\cdot) $ is the  solution of ODE \eqref{Equation_epsilon_simplified_equation}, with    $\up(0) = \lim_{k_m \to \infty} \up_{k_m}$ as the initial condition. 
\item [ii)] Additionally assume  \textbf{A.} Then the sequence converges, $\up_k \to {\cal A}$ as $k \to \infty$ with probability at least $\varrho$.
\end{enumerate}
\end{thm}
{\bf Proof} is in Appendix. 
\eop

{\bf Remarks:} i) The above theorem includes Theorem \ref{Thm_RandomC_conv} as a special case (obtained when $E[{\cal L}] = 0$). 
ii) The  above theorem   also presents a finite horizon approximation result in  part(i), which hence justifies the approximation used by equation \eqref{Eqn_ode_approx}  of Section \ref{Sec_finance_replicatordynamics}.
iii) In the finite horizon approximation,  we initialized the ODE solution with the limit of the random trajectory, $\lim_{k_m \to \infty } \Theta_{k_m}$.  One can derive a much better approximation when the ODE solutions are continuous in initial conditions  (e.g., as in \cite{perko}), by initializing  $\Theta(0) = \Theta_{k_m}$; this requires a small obvious addition to the proof provided in Appendix. For many case studies (see ODE solution \eqref{Eqn_ode_sol_with_limiteed_default} given below) we definitely have continuity with respect to initial conditions. We in fact use this approximation  for all  the numerical examples  provided in Section \ref{Section_numerical_observations}.

\noindent{\bf ODE Solution and analysis:} 
We derive the solution of the coupled ODE \eqref{Equation_epsilon_simplified_equation} in the following. Towards this, we define four intervals and first obtain the required solutions when it (actually $\epsilon(t)$) is confined to the intervals: 
\begin{eqnarray}
\label{Eqn_interval_ode}
    {\cal I}_0 = \left [0, \bar{\epsilon_1} \right], \  {\cal I}_1 = [ \bar{\epsilon_1}, \bar{\epsilon}),  \ 
    {\cal I}_2= \left [\bar{\epsilon}, 1+\frac{E[\mathcal{L}]}{\beta} \right ), \mbox{ and, } \ {\cal I}_3  =  \left (1+\frac{E[\mathcal{L}]}{\beta}, 1 \right ) \hspace{-1mm}. \hspace{4mm}
\end{eqnarray}
where,  $\bar{\epsilon}_1$, and  $\bar{\epsilon}$, are defined in  Lemma \ref{Lemma_mono_PD} and Theorem \ref{Theorem_financenetwork}   respectively. Also let ${\cal D}_\psi :=[0, 1+{\bar \Nw}]$ and consider any initial condition $\Theta (t_0) = \Theta_{t_0} \in   {\cal I}_i \times {\cal D}_\psi$ for some $i \in \{0, \cdots, 3\}$ (ODE-dynamics can start at arbitrary initial time $t_0$). 
It is easy to verify\footnote{\label{footsolution}
i) As seen from proofs in Appendix  the random trajectory $\psi(t)$ is upper bounded by $1+{\bar {\cal N}}$, hence sufficient to consider domains with ${\cal D}_\psi$;\\ ii) it is easy to verify that   $\psi(t) = (\psi_{t_0} -a_i) e^{-t} +  a_i$  when $\epsilon(t)$ starts in  interval ${\cal I}_i$, while the solution $\epsilon(t)$ can be derived using elementary calculus-based steps like
$$
\int_{\epsilon_{t_0}}^{\epsilon(t)} \frac{d \epsilon} {\kappa \epsilon (1-\epsilon) + \epsilon E[{\cal L}] } = \int_{t_0}^t  \frac{ds}{\psi_s}.
$$
iii) The sign of $\kappa + E[{\cal L}] - \kappa \epsilon  $  remains the same for any  $\epsilon \in {\cal I}_i$, and hence  
$$\log \left ( \frac{ \mid \kappa + E[{\cal L}] - \kappa \epsilon(t)    \mid  } { \mid \kappa + E[{\cal L}] - \kappa \epsilon_{t_0}  \mid } \right )  = \log \left (\frac{\kappa + E[{\cal L}] - \kappa \epsilon(t) }{ \kappa + E[{\cal L}] - \kappa  \epsilon(t_0) } \right ).$$ 
iv) Observe that the resultant solution $\epsilon(t)$ is strictly monotone as long as $\epsilon(t)$  is confined in  ${\cal I}_i$. 
} 
that we have a unique global solution of ODE \eqref{Equation_epsilon_simplified_equation} when $\epsilon(t)$ is started (and as long as it is confined)  in the interval ${\cal I}_i$.  Further, the  ODE solution for any $t_0 \le t \le \tau$, with $ \tau := \inf \{s: \epsilon(s) \notin {\cal I}_i \}$,  is given  (refer footnote \ref{footsolution}) by:
\begin{eqnarray}
\label{Eqn_ode_sol_with_limiteed_default}
\psi(t) &=& (\psi_{t_0} -a_i) e^{-t} +  a_i,  \\
\epsilon(t) &=& \frac{ \mu_i \epsilon_{t_0}  h_i(t)  } {\mu_i-\epsilon_{t_0} + \epsilon_{t_0} h_i(t) }, \mbox{ with , }  h_i(t) \ = \  \left ( \frac{a_ie^{t}+ \psi_{t_0} - a_i }{a_ie^{t_0}+  \psi_{t_0} - a_i } \right )^{Q_i},  \mbox{ where, } \nonumber \\
&&\begin{array}{lllll}
\mu_0 = 1,     &   \mu_1 = \frac{ \beta (1-2\delta)+ E[\mathcal{L}]}{ \beta (1-2\delta)},    
     & \mu_2= \mu_3= \frac{\beta + E[\mathcal{L}]}{\beta }, \\
 Q_0 = \frac{\beta(1-2\delta)}{E\Nw},  \hspace{3mm}   & Q_{1}  =    \frac{ \beta (1-2\delta) +E[\mathcal{L}]}{E[\Nw]-E[\mathcal{L}]},   &  Q_{2} \ =  Q_3 = \   \frac{\beta +E[\mathcal{L}]}{E[\Nw]-E[\mathcal{L}]}, \\
 a_0 = E[\Nw], \mbox{ and, } & &   a_1 = a_2 = a_3  = E[\Nw]-E[\mathcal{L}].
\end{array} \nonumber 
\end{eqnarray}
For simpler discussions assume   $\mu_i \ne 0$ for all $i$ and ${\bar \epsilon}_1 > 0$. From solution \eqref{Eqn_ode_sol_with_limiteed_default}, we immediately have  the following:
\begin{lemma} 
\label{Lemma_ODE_Analysiswithdefault}
Consider any initial condition  with {\small $\epsilon_{0}:=\epsilon_{t_0}  \in (0,1)$.  Also  assume,   $ \epsilon_{0} \notin  \{ 1+ \nicefrac{E[{\cal L}]}{\beta (1-2\delta)},1+ \nicefrac{E[{\cal L}]}{\beta} \}  .$} 
\begin{enumerate}
\item [a)]  If $Q_2 > 0$  then $\epsilon(t) \to 1$ $\forall$ $\epsilon_0 \ge \bar{\epsilon}$ as $t \to \infty$, and also $\nicefrac{E{\cal [L]}}{\beta} > 1.$
\item [b)]   If $Q_2  <0$  and $\epsilon_0 > 1+\frac{E[\mathcal{L}]}{\beta}$ then $\epsilon(t)  \to 1$  as $t \to \infty$.
\item [c)]  If $Q_2  <0$  and $\epsilon_0 < 1+\frac{E[\mathcal{L}]}{\beta}$ then $\epsilon(t) $ is a decreasing function of $t$ until   $\epsilon(t) \ge \bar{\epsilon}$. 
\item [d)] The function $\epsilon(t) $ is an increasing  (resp., decreasing) function of $t$  as long as  $\epsilon(t)  \in {\cal I}_2$, if  $Q_1$ is positive (resp., negative).  

\item [e)]  The function $\epsilon(t)$ is an increasing  (resp., decreasing) function of $t$  as long as  $\epsilon(t) \in {\cal I}_1$, if  $\beta (1-2\delta)$ is positive (resp., negative). 
\end{enumerate}
\end{lemma}
\noindent
{\bf Proof:} The function $h_i(.)$ defined in \eqref{Eqn_ode_sol_with_limiteed_default}, for any $i$ is strictly monotone  (decreasing or increasing based on $Q_i$ and note $a_i > 0$) in time $t$, and the mapping $x \mapsto \nicefrac{\mu_i \epsilon_0 x}{(\mu_i - \epsilon_0 + \epsilon_0 x)}$ is also strictly monotone when $\mu_i \ne 0$. The rest of the details can be easily verified.  \eop

The above lemma immediately implies that $\epsilon^* \in \{0, 1\}$ characterize the potential attractors of the ODE (depending upon the parameters) and one can again have a mixed ESS at $\epsilon^* = {\bar \epsilon}$, as in Section \ref{Sec_finance_replicatordynamics}.  One can also anticipate that assumption {\bf A} would be satisfied, with  DoA equal to $[0, 1] \times {\cal D}_\psi$, and  can easily observe that the trajectory visits such a DoA infinitely often with probability one. To make precise these observations, we consider an important case study (there are too many possibilities as seen from the Lemma \ref{Lemma_ODE_Analysiswithdefault}, hence can make precise statements only after considering particular case studies). 

In financial markets   typically the probability of downward movement is small, hence it is reasonable to consider a case study with $\delta > 1/2$ (analysis can be derived for the case with $\delta \le 1/2$ exactly as below). 

Suppose now $\beta(1-2\delta) < 0$, then $\beta > 0$. This implies $Q_2 > 0$.   For this case,  from Lemma \ref{Lemma_ODE_Analysiswithdefault} and the   solution \eqref{Eqn_ode_sol_with_limiteed_default}, one can easily verify that: \\
a)  we have unique global solution for all $t$  (either the solution $\epsilon(t)$ keeps increasing or keeps decreasing in a series of intervals given in \eqref{Eqn_interval_ode},  based on initial conditions); \\
b) if $Q_1>0$ then 
$(\epsilon^*, \psi^*) = (1, E[\Nw])$ is one of the attractors and its
DoA, ${\cal D}_1 =  [{\bar \epsilon}_1, 1] \times {\cal D}_\psi$; \\
c) if $Q_1 < 0$ the  DoA for the same attractor is, ${\cal D}_1=[{\bar \epsilon}, 1]\times {\cal D}_\psi$; \\
d) we have a second attractor, $ (0, E[\Nw])$   with  DoA ${\cal D}_0 =  [0, 1]\times {\cal D}_\psi -  {\cal D}_1$; and   \\
e) thus the assumption {\bf A} is satisfied with attractors and combined DoA   as, 
$${\cal A} = \{ (0, E[\Nw]), (1, E[\Nw])\},    \mbox{ and,  } {\cal D} = [0, 1] \times {\cal D}_\psi \mbox{ and with } \varrho = 1.$$

Consider $\beta(1-2\delta) > 0$
and $\beta < 0$, and consider two further sub-cases i) with  $\beta + E[{\cal L}] > 0$ (i.e., when $Q_2 > 0$); or ii)  with $\beta + E[{\cal L}] < 0$ ($Q_2 < 0$ and $1+ \nicefrac{E[{\cal L}]}{\beta } < {\bar \epsilon}$).   Again from   Lemma \ref{Lemma_ODE_Analysiswithdefault} and the ODE solution \eqref{Eqn_ode_sol_with_limiteed_default},   
set of attractors and DoA are   ${\cal A} = \{(1, E[\Nw])\}$ with DoA $[0, 1]  \times {\cal D}_\psi$.

The leftover case with $\delta  > 1/2$, i.e., when $\beta(1-2\delta) > 0$, 
  $\beta + E[{\cal L}]  < 0$ and $1+ \nicefrac{E[{\cal L}]}{\beta } > {\bar \epsilon}$, will have two $\epsilon
  $-limits $\{{\bar \epsilon}, 1\}$ depending upon the initial conditions (see Table \ref{Table_conditionsofattractor} for a complete description of the attractors).  The dynamics converge to mixed limit ${\bar \epsilon}$  (when initial condition $\epsilon_{0} < 1+ \nicefrac{E[{\cal L}]}{\beta } $) as in Section \ref{Sec_finance_replicatordynamics}; the rest of the  details of this ESS  are exactly similar.  
  \begin{table}[htbp!]
\begin{center}
\begin{tabular}{|c|c|c|}
\hline
Parameters   & Attractors with $E[{\cal L}] > 0$ &  Attractors with $E[{\cal L}] = 0$ \\  
 & Some Defaulters departing  &  \\ \hline
$\beta > 0$ &  $ (0, \ E[\Nw])$  &  $ (0, \  E[\Nw])$ \\ 
& and $(1, \ E[\Nw])$ & and $(1, \ E[\Nw])$  \\ \hline
$\beta <  0$ and &  &  \\ 
  $\beta+E[{\cal L}] > \beta {\bar \epsilon}$
 & $(1, \  E[\Nw])$ &   $(\bar{\epsilon}, \  E[\Nw])$ \\ \hline
 $\beta <  0$ and &   $({\bar \epsilon}, \    E[\Nw]-E[{\cal L}])  $  &   \\ 
   $\beta+E[{\cal L}] < \beta {\bar \epsilon}$
 & and  $(1, \  E[\Nw])$ &   $(\bar{\epsilon}, \  E[\Nw])$ \\ \hline 
  
\end{tabular}
\end{center}
\vspace{2mm}
 \caption{ODE-attractors with and without defaulters stopping investments (with $\delta > 1/2$).}\label{Table_conditionsofattractor}
\end{table}
One can summarize the set of attractors (or the limits of financial dynamics) for the case with $\delta > 1/2$ in Table~\ref{Table_conditionsofattractor}; we also compare this case with the case in which the defaulters would not stop   investments in further rounds. As seen from the table, the case with perfect information $\beta  > 0$ has a similar asymptotic outcome for both cases (the DoAs for individual attractors have changed, as seen above and in Corollary \ref{Lemma_ODE_Analysis}.c).  More interestingly, in the case with imperfect information, i.e., with $\beta < 0$, when $E[{\cal L}] >0$, in many cases, the system reaches a state with only risk-free agents; the counterpart with $E{\cal [L}]=0$ reaches the intermediate mixed ESS, unless ${\bar \epsilon} \in \{0,1\}$  (see Corollary \ref{Lemma_ODE_Analysis}.c).  However, whatever may be the case, the systemic risk regime is again avoided even for the case with  $E[{\cal L}] >0$.
 In fact, with $E[{\cal L}] >0$, the system even avoids the regimes where there is non-zero number of defaults; for example, in   the second row of the Table \ref{Table_conditionsofattractor} the system converges to $\epsilon^* = 1$ while the same with $E[{\cal L}] = 0$ converges to ${\bar \epsilon}$; of course in the third row such a situation is only avoided  partially (both $1$ and ${\bar \epsilon}$ are attractors). 

\section{Numerical Observations}
\label{Section_numerical_observations}
This section reinforces our theoretical findings using the estimates obtained from  Monte-Carlo (MC) simulations. For each run of the simulation: a) a sequence of binomial shocks, sequence of random connection,  sequence of users joining, random sampling, etc. are realized;  b) at each round of each run, the clearing vector (fixed point equations) are solved using iterative methods to obtain the clearing vector; c) using the clearing vectors we estimate  the random returns of all the agents of that round; and d)  at the end of each round the  comparisons of random samples (agents/strategies) are made using the 
(estimated)   returns of that round to continue the dynamics as in  Section \ref{Sec_finance_replicatordynamics}. We have summarised the procedure in Algorithm \ref{Algo_fixedpoint}, provided in Appendix.

In Table \ref{Table_MC _basedwithrandom addition_random}  we consider the first example with asymptotic limits. 
Each run of the simulation is considered  for  $T=1000$ number of rounds and with initial population size $n_0 = 300$. We set   $E[\Nw_t] =1$ and  $E[\Sw_t] =10$ while the rest of the details are in the table itself.   The results are tabulated   for various configurations.
The configurations are specified in the first two columns,   $\epsilon^{*}_{Th}$ is the theoretical limit obtained using ODE \eqref{Eqn_ODE_epsilon}, and  $\epsilon^{*}_{MC}$ is the MC based estimate. We also included ${\bar \epsilon}$ of Theorem \ref{Theorem_financenetwork}.  
The first observation is that the   theoretical ODE-limits  well  match the MC-based limits.
 \begin{table}[htb!]
\centering
\begin{tabular}{|l|c|c|c|c|}
\hline
Configuration ($b_s$, $b_n$, $\alpha$, $\delta$) & $\epsilon_0$ & $\epsilon^{*}_{Th}$ & $\epsilon^{*}_{MC}$ &  $\bar{\epsilon}$ \\ \hline
(0.9, 0.9, 0.95, 0.85)   & 0.85 &  1&  0.9866 & 0.8350 \\ \hline
(0.9, 0.9, 0.95, 0.85)   & 0.75 &  0.0004 &   0.0008 & 0.8350 \\ \hline
(0.4, 0.4, 0.95, 0.85)   & 0.8 &  0.8350 &   0.8367  & 0.8350 \\ \hline
\hline
(0.9, 0.9, 0.95, 0.45)   & 0.6 &  1 &   0.9985  & 0.2233 \\ \hline
(0.15, 0.15, 0.95, 0.45)   & 0.2 &  0.0749 &    0.0649 & 0.2233 \\ \hline
\end{tabular}
\vspace{2mm}
\caption{MC estimates versus ODE-limits:  $v= 20, w=70$, $ (u, r_s, r_b, d) = (0.15, 0.1, 0.11, -0.6)$.}\label{Table_MC _basedwithrandom addition_random}
\end{table}

The first two rows consider a configuration   with $\beta > 0$ and ${\bar 
\epsilon} < 1$. Results well corroborate with Corollary \ref{Lemma_ODE_Analysis}.(a); when initial $\epsilon_0 = 0.75 < {\bar \epsilon}$, the random trajectory converges to 0; the limit is 1 if $\epsilon_0 = 0.85$. 
The third row considers the same configuration, except for $b_n=b_s = 0.4$. With this  partial information     ($\beta < 0$),   we have the conjectured mixed ESS:  
the trajectory wanders around $\bar{\epsilon}$. In fact, we noticed similar  convergence around $\bar{\epsilon}$ for many  other cases. 

In the last two rows, a case with  a large probability of economic shocks is considered ($\delta = 0.45$). Depending upon the quality of   information ($\beta <0$  or $\beta >0$), the dynamics converge to a pure limit  ($\epsilon_t \to  1$ or $\epsilon_t \to  0$ respectively). This reinforces the results of Corollary \ref{Lemma_ODE_Analysis}. (c).

 \begin{figure}[ht]
 \vspace{-6mm}
     \centering
     \includegraphics[scale=0.3]{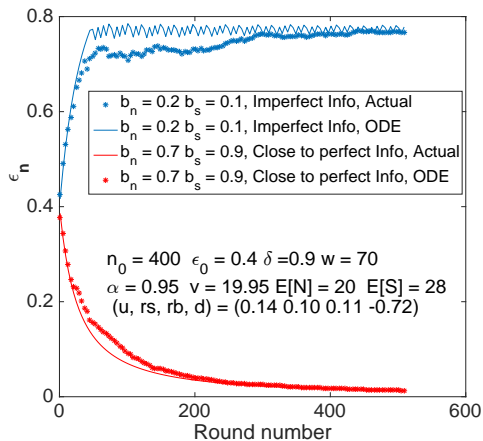}
     \vspace{-10mm}
          \caption{Imperfect ($\beta < 0$) versus Perfect ($\beta > 0$) Information}
    
     \label{fig:perfect_imperfect}
     \vspace{-4mm}
 \end{figure}
\begin{figure}[ht]
\vspace{-8mm}
     \centering
     \includegraphics[scale=0.3]{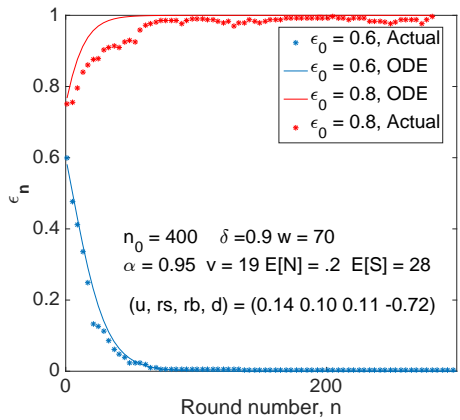}
\vspace{-12mm}
\caption{Predominant switching with   E[\Sw]=28 and  $E[\Nw] =.2$}

     \label{fig:majorlyswitch}
          \vspace{-3mm}
 \end{figure}


            
            
              

  In figures \ref{fig:perfect_imperfect}-\ref{fig:majorlyswitch} we illustrate the accuracy of finite horizon approximation by plotting ODE-based  trajectories   \eqref{Eqn_ode_approx}  and the corresponding MC trajectories   for four configurations. 
 The two curves close-in as the number of rounds increases. This is true even for  the case in Figure  \ref{fig:perfect_imperfect}, where we have  $w(1+d) < v$, which is not covered by theory.

 
 In the  Figure \ref{fig:perfect_imperfect}, we have a case  
 with a comparable fraction of new agents  ($E[\Nw]=20$ and $E[\Sw]=28$).  Here we compared the scenarios with and without 
 perfect information.  
 When $\beta >0$ (red curves) the dynamics converge towards $\epsilon^{*} =0$, i.e.,  all the agents settle for the `risky' strategy. 
 With erroneous information, i.e., with  $\beta  < 0$,  the dynamics (blue curves) settle to  mixed fraction ${\bar \epsilon}.$ In both the scenarios, dynamics  avoid the systemic risk   regime ($P_d =1 $   only for $\epsilon > {\bar \epsilon}_2$ and ${\bar \epsilon} \le {\bar \epsilon}_2$, see Lemmas \ref{Lemma_threshold of q_eps}, \ref{Lemma_mono_PD}). 
 
 In Figure \ref{fig:majorlyswitch} we consider the scenarios with perfect information and predominant switching; we observe that the dynamics settle to a pure ESS, i.e., $\epsilon^{*}=0$  or  $\epsilon^{*}=1$, depending upon  $\epsilon_0$.

\begin{figure}[htbp!]
 \vspace{-16mm}
     \centering
     \includegraphics[scale=0.4]{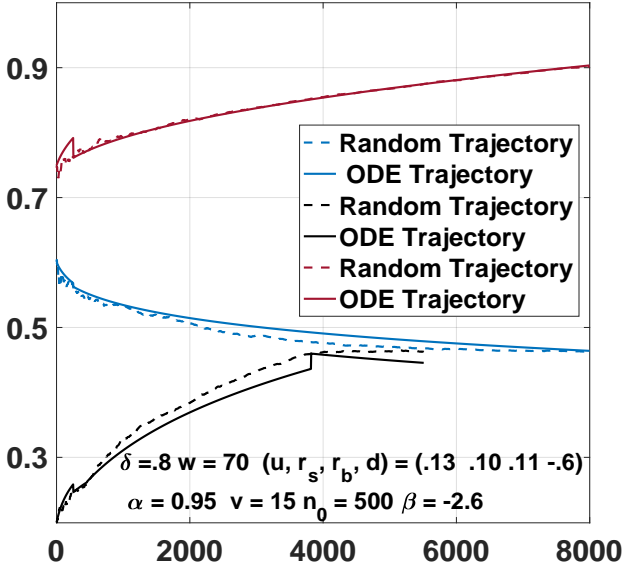}
     \vspace{-13mm}
    \caption{ With imperfect information ($\beta<0, b_s=b_n=0.4$) and  $E[\mathcal{L}]=0.84$. The ODE solution restarted after 250 rounds.} 
    \label{Fig_DGAA_AllInit}
    \vspace{-4mm}
 \end{figure}

 \textbf{Defaulters stop investing:} 
We now consider an MC simulations study related to Section \ref{Section_defaultersleave}; in each simulation run,  some  of the defaulted agents leave the network; some random number  of agents join, and a random number of them  switch the strategies. The rest of the details are as before.
We have used the following common set of parameters for the case studies in  Table \ref{Table_with_default_attaractor_perfect}-\ref{Table_with_default_attaractor_imperfect} and Figure \ref{Fig_DGAA_AllInit}: $u=0.13$, $d=-0.6$, $r_s= 0.1$, $r_b= 0.11$, $w= 70$, $\delta=0.8$, $\alpha=0.95$, $n_0=500$, $v=15$, $T= 4000$,  $E[\Nw] =7$, $E[\Sw] =6$.

In  Tables \ref{Table_with_default_attaractor_perfect}-\ref{Table_with_default_attaractor_imperfect}, we compared the asymptotic  limits   when  some of the defaulters leave the network versus when they do not. Moreover, we consider  a comparison with and without perfect  information. In all the tables,   the theoretical estimates (given in Table \ref{Table_conditionsofattractor}) and MC estimates   match closely;  we again observe the differences between the two case studies exactly as depicted by the theoretical results summarized in Table \ref{Table_conditionsofattractor}.

In Table \ref{Table_with_default_attaractor_perfect}  with perfect information,   the first row shows that the asymptotic limits are different with and without the defaulters leaving the system. The convergence depends on the initial proportion, $\epsilon_0$. 
In  the first row of Table \ref{Table_with_default_attaractor_imperfect}, with imperfect information and high initial proportion  $\epsilon_0 \in \{0.4,0.8\}$, the financial dynamics settle to a configuration with  all less-risky agents,  when some defaulters stop investing.  On the other hand,  with all the defaulters continuing the investments in further rounds, the dynamics converge to a mixed limit.

\begin{table}[htbp!]
\begin{center}
\begin{tabular}{|c|c|c|}
\hline
 $\epsilon_0$&  ($\epsilon^{*}_{Th}, \epsilon^{*}_{MC}, E[{\cal L}]$) &  $(\epsilon^{*}_{Th}, \epsilon^{*}_{MC},  E[{\cal L}]=0)$\\ \hline
 0.4&  (1, 1, 5.6) & (0, 0.0429, 0)  \\ \hline
 0.3 & (0, 0.061, 2.1) &  (0, 0.0295, 0) \\ \hline
 \end{tabular}
\end{center}
\vspace{4mm}
\caption{Perfect information ($\beta=7.8 > 0$, $b_s=b_n=0.8$). Further,  $\bar{\epsilon}=0.4597$,   $\bar{\epsilon}_1= 0.2616$.  }\label{Table_with_default_attaractor_perfect}
\end{table}
\begin{table}[htbp!]
\begin{center}
\begin{tabular}{|c|c|c|}
\hline
 $\epsilon_0$&  ($\epsilon^{*}_{Th}, \epsilon^{*}_{MC}, E[{\cal L}]$) &  $(\epsilon^{*}_{Th}, \epsilon^{*}_{MC},  E[{\cal L}]=0)$\\ \hline
 0.4 &  (1, 1, 1.75) & (0.4597, 0.4521, 0)  \\ \hline
 0.8 & (1, 1, 1) &  (0.4597, 0.4803, 0) \\ \hline
 0.5 & (0.4597, 0.4627, 7) &  (0.4597, 0.4589, 0) \\ \hline
 \end{tabular}
\end{center}
\vspace{5mm}
\caption{Imperfect information (with $\beta =-2.6< 0$, $b_s=b_n=0.3$). Further,  $\bar{\epsilon}=0.4597$, $\bar{\epsilon}_1= 0.2616$.  }\label{Table_with_default_attaractor_imperfect}
\end{table}

In Figure \ref{Fig_DGAA_AllInit}, we demonstrate the accuracy of finite horizon approximation by plotting ODE-based trajectories and the corresponding MC trajectories for the same configurations, but with three different initial conditions. In all the cases, the two trajectories  close in as the number of rounds increases. 
Further, the asymptotic outcome matches with the theoretical result as in Table \ref{Table_conditionsofattractor}.

Another important observation is that the dynamics avoid the systemic risk regime (where all  the $G_2$ agents default), even when a moderate number of defaulters stop investing.

An important concluding remark is regarding the impact of rational behaviour of the agents, because of which systemic risk regime is  avoided in   all the scenarios considered in the paper. Our analysis can also be used  to show that the system actually leads to a systemic risk regime if the agents blindly  adapt the strategies; for example, in the scenarios that have $1$ or ${\bar \epsilon}  $ as ESS, if instead all the agents blindly adapt the  risky strategy, then by equation 
\eqref{Eqn_mono_Pd} of Lemma \ref{Lemma_mono_PD} in Appendix,  we have  $P_d = 1$ (i.e., all the agents would have defaulted). 
Thus the rational agents avoid systemic regime (ESS or limit is $0$ or ${\bar \epsilon} > 0$ respectively implying zero or only a fraction of agents default), while with blind agents (adapting risky strategy)  all the agents default leading to systemic risk regime.

 \section{Conclusions}
We consider a financial network wherein the agents are interconnected via liability edges/connections. There are two types of agents; one group lends to others and invests in risk-free projects, while the second group borrows/lends and invests in risky ventures. 
These agents try to adapt their strategies based on their experiences and observations. Some new agents may also join the network. Some existing agents may leave the network.  Thus we have a sequence of random networks evolving with time, where the strategic agents are changing their connections locally to improve their returns. 
We analyze such evolution and establish the emergence of  evolutionary stable strategies. Towards this,   we reduced the analysis of the evolution of this complex time-varying financial system to that of an  appropriate ordinary differential equation (ODE);   some recent results  
helped reduce the large dimensional random fixed point equations to simpler representations, which are instrumental in deriving the ODE.  

Using the attractors of the resulting ODE, we showed that with perfect information, the replicator dynamics converges to one of the two pure evolutionary stable strategies, i.e., configurations with all `risky' or all `less risky' agents. With imperfect information, dynamics can settle to a mixed limit. In all the cases, the dynamics averted  the `systemic risk event', where most agents default; in case all agents blindly adapt risky strategy it can lead to `systemic risk event'. 
We also performed Monte-Carlo simulations to reaffirm the theoretical findings.

\section*{Author contributions}
All authors contributed to the study, conception and design. The first draft of the manuscript was written by Indrajit Saha, while,  Veeraruna Kavitha helped in improving the manuscript.   All authors read and approved the final manuscript.

\section*{Data availability}
Data sharing is not applicable to this article as no datasets were generated or analysed during the current study. We generated synthetic data which is included in  the article itself. 

\section{Appendix}
\label{Section_appendix}

\begin{algorithm}[htbp!]
\caption{Financial Replicator Dynamics}
\begin{algorithmic}[1]
    \State  Inputs:  $n_0$, $\epsilon_0$, $w$, $v$, $\delta$, $\alpha$, $b_s$, $b_n$, $r_s$, $r_b$, $u$, $d$, $p_{ss}$, $T$ 
    \State  {\bf Finance dynamics (Rounds)} For each round $t= 0, 1, \cdots ,T$:

    \begin{itemize}
        
   \item Generate $K_i$, $\{I_{j,i}\}$ to create an instance of the network corresponding to $(n_t, \epsilon_t)$.

    \item Initialize clearing vector $  X^0_i  =  y$, i.e., for step $k=0$ (of round $t$) and for all $i\le n_t$ where $y = y(\epsilon_t)$ is given by  equation \eqref{Eqn_liability}.
        \State {\bf Compute Clearing Vector for round $t$:} Run the sub-algorithm  to compute the fixed-point (clearing vector),  for steps $k = 0, 1, \cdots, $  till convergence (verified  using $(\delta_0, l)$).   
     \begin{itemize}
            \item  update: $X_i^{k+1} =  X_i^{k} - \gamma_k(f_i({\bf X}^k)- X_i^{k})$ 
            
            \item if $\sum_{i=1}^{n}\lvert X_i^{s+1} - X_i^{s}\lvert< n\delta_0 $, for all $s = k-l, k-l+1, \cdots, k$
            
            \item  algorithm  converged at $k^* = k$ and end  
            \end{itemize}
            Here $f_i({\bf X}^k):= \min \left  \{  \bigg( K_i+ \sum_{j\in G_2} X_j^k\frac{L_{ji}} {y} - v \bigg )^+, \    y \right \}$.
            \item  {\bf Estimate the performance metrics of $G_2$,} using clearing vector:
            \begin{itemize}
           \item For each $i$, the amount cleared by    agent $i$ =
$                X_i^{k^*}  
               $
                
                \item             Default probability $P_d = \frac{1}{n_t}\sum_{i\le n_t} 1_{\{X_i^{k^*} < y\}} $, where

                \item Returns of all agents, for agent $i$ (see equation \eqref{Retun for G_2}),  
                $$\hspace{-3mm}
                R^2_i  = \bigg( K_i+  {\bar X}_i^{k^*} - v  - y \bigg )^+  \mbox{ if in group $G_2$  and  } R^1_i = \left  ( w\epsilon(1+r_s)+ {\bar X}_i^{k^*}  -v \right )^+  $$
                 if in group $G_1$, where 
                $$
                {\bar X}_i^{k^*} = \sum_{j\in G_2} X_j^{k^*}\frac{L_{ji}} {y}.
                $$
            \end{itemize}

            \item  {\bf Switching entries: Generate the    random variable} $\xi_t$.  For each $i = 1, 2, \cdots, \xi_t $
            
            \begin{itemize}
                \item  Draw two random agents of  the previous round ($t-1$)
                \item  Compare the  random returns of the sampled agents and,
                \item Update the switching entity to the correct group  based on sampled returns.
            \end{itemize}

            \item  {\bf New entries: Generate the   random variable} $\mathcal{N}_t$. For each $i = 1, 2, \cdots, \mathcal{N}_t$ 
            \begin{itemize}
                \item  Draw two random samples  from the previous round $(t-1)$
                \item  Compare the  random returns of the sampled agents and,
                \item Add the entity  to correct group based on the sampled returns.
            \end{itemize}
            \item Update $n_1(t)$, $n_2 (t)$ using the above two modifications to generate $n_1(t+1), n_2 (t+1)$.

             \end{itemize}

            \State {\bf Update:}  $t \leftarrow (t+1)$  and  $\epsilon_t \leftarrow n_1(t)/n(t)$
         \State end
    \end{algorithmic}
\label{Algo_fixedpoint}
\end{algorithm}

\subsection{Details of Asymptotic Approximation}
\label{sec_asym_approx}
  The clearing vector  \eqref{Clearing vector}, 
 defined using equations \eqref{Eqn_liability}-\eqref{Clearing vector},
 can be viewed as the solution of   random fixed point equations, which depend upon the realizations of the economic shocks  $\{K_i\}_{i \in {G}_2}$ to the network. We obtain an approximate clearing vector by  applying the single group results   of  \cite[Corollary 1 and  Subsection 4.2]{saha2021random} only to group $G_2$. Towards this
 we consider  a fictitious big node (like in \cite{saha2021random}) and from each node $j \in G_2 $ there is a dedicated  fraction $(1-c_\epsilon) $ with $ c_\epsilon: = \frac{\alpha +\alpha \epsilon}{\alpha +\epsilon}$  directed towards the fictitious node. This financial system is exactly similar to  the graphical model described by \cite[equations (1)-(6) and (40)-(41)]{saha2021random}, after the   following mapping   details:
\begin{eqnarray}
 G_i &=& K_i, \  \eta_i^{bs} =0  \  (a.s.), \ W_{j,i}=\frac{L_{ji}}{y}= \frac{I_{ji}c_\epsilon}{np_{ss}(1-\epsilon)}, \  p_{c_1} =p_{c_2}= 0, \nonumber \\ 
p_1 = p_2&=& p_{ss},  \mbox{ and }  \ p^{sb}_1= p^{sb}_2 = 1- c_\epsilon. 
\label{Eqn_ess_mappings}
\end{eqnarray} With the above mapping details the required assumptions  of \cite{saha2021random} are satisfied:  assumption {\bf B.1} is immediately satisfied (see  \eqref{Clearing vector}),  assumption {\bf B.3} is  satisfied with $\sigma = 1$ and  with any $0 \le \varsigma < 1$ (as the fixed point equations do not depend upon $x_b$). 
The weight factors are  as in  \cite[  Subsection 4.2]{saha2021random} and hence assumption {\bf B.2} is not required. Finally, the assumption  {\bf B.4} is satisfied with $\rho=1$. Hence by \cite[Corollary 1]{saha2021random} (as $0 < c_\epsilon < 1$ with $\epsilon \in (0,1)$),  the solution of the random  fixed point equations \eqref{Clearing vector} can be  approximated using that  corresponding to the limit system given in   \cite[Corollary 1]{saha2021random}. 
 Thus  we have the convergence provided\footnote{One can partially justify similar approximation for systems with $\epsilon \in \{0, 1\}$   using \cite[Theorem 1 and  Subsection 4.2]{saha2021random}.}
 in equations \eqref{Eqn_aggregate_limitof individualgroup}-\eqref{Eqn_clearingvector_limit} 
 for any 
$\epsilon \in (0,1)$ for $G_2$ (as the network size increases to infinity). This in turn provides
the convergence  results for $G_1$.

 \textbf{Network with only risky agents ($\epsilon=0$):}     The total amount lend by any agent to its neighbours equals $\approx w \alpha / (1-\alpha)$, where the approximation is again accurate at limit. In a similar way the total amount borrowed by any agent also equals $\approx w \alpha / (1-\alpha)$. Thus
any agent invests $\approx w$ (their initial wealth) in risky assets. In this case  the limit aggregate clearing vector \eqref{Eqn_aggregate}  reduces to the following:
\begin{equation}
\label{Equation_barx_ateps=0}
 {\bar x}^{ \infty} =E \bigg[\min \left  \{  \bigg( K_i+  {\bar x}^{ \infty}  - v\bigg)^+,  y  \right \}\bigg].
\end{equation}
When $w(1+d) \ge v$ it is easy to observe that $\bar{x}^{\infty}= y$ is the unique solution of the  above equation. Hence $P_d=0$. Therefore $R_i^{2}=K_i -v$ at limit for any $i$. For this case   \cite[Corollary 1]{saha2021random} is not applicable, however   \cite[Theorem 1]{saha2021random} (applied to single group as in
 \cite[subsection 4.2 1]{saha2021random}, partially justifies the above approximation.

 \textbf{Network with only  less risky  agents ($\epsilon=1$):} On the other hand with $\epsilon= 1$, all are  less risky agents and  they invest completely in risk-free assets. Thus the return of any agent $i\in G_1$ equals, $R_i^{1}=w(1+r_s) - v$. 
 
 \medskip
 
 \subsection{Proofs}

\noindent \textbf{Proof of Lemma \ref{Lem_Average_clearing}:}  We consider the following  scenarios with $v< k_d$. The average clearing vector  for the group $G_2$ agents  satisfies (see \eqref{Eqn_aggregate}):
\begin{eqnarray}
\label{Eqn_simplification_clearing}
{\bar x}^{ \infty} &=& \min \left  \{  \bigg( k_d+  c_\epsilon  {\bar x}^{ \infty}  - v\bigg)^+,  y  \right \}(1-\delta) \nonumber \\ 
&& + \min \left  \{  \bigg( k_u+  c_\epsilon  {\bar x}^{ \infty}  - v\bigg)^+,  y  \right \} \delta.
\end{eqnarray}
\textbf{Case 1:} First consider the case when downward shock can be absorbed i.e., default
probability is $P_d= 0$. If we have $k_d +yc_\epsilon -v \ge y$ then the average clearing vector  ${\bar x}^{ \infty} = y \delta + y(1-\delta) =y$, and the above condition simplifies to the bound:
  \begin{equation}
  k_d-v +yc_{\epsilon} \ge y \implies c_{\epsilon} \ge \frac{y-\underline{w}}{y}.\nonumber
  \end{equation}
 \textbf{Case 2:} Consider the case in which only the agents that receive shock will default,  i.e., when $P_d =1-\delta$. The corresponding average clearing vector equals: 
 
 \vspace{-2mm}
 {\small
 \begin{eqnarray*}
 {\bar x}^{ \infty} = y\delta+ (\underline{w} +c_\epsilon {\bar x}^{ \infty})(1-\delta)  \mbox{ and satisfies }
  k_d- v +c_{\epsilon} {\bar x}^{ \infty} < y, \mbox{ }
  k_u- v +c_{\epsilon} {\bar x}^{ \infty} \ge y.
 \end{eqnarray*}}
 In this case the average clearing vector reduces to 
   $
   {\bar x}^{ \infty} = \frac{y\delta+ \underline{w}(1-\delta) }{1-c_\epsilon(1-\delta) },
   $ 
   and using the same in the bounds we have:
\begin{equation}
\frac{y-\overline{w}}{y-(1-\delta)(\overline{w}-\underline{w})} \le
c_\epsilon <\frac{y-\underline{w}}{y}.\nonumber
\end{equation}
\textbf{Case 3 (Systemic-risk regime):}
Consider the case in which  all the agents default i.e., when $P_d
= 1$. In this we first calculate ${\bar x}^{ \infty}$ which is obtained by solving following fixed point equation:
If we have $k_u- v+ c_\epsilon {\bar x}^{ \infty} < y$ then from  \eqref{Eqn_simplification_clearing} the average clearing vector reduces to:

\vspace{-4mm}  
{\small
  \begin{eqnarray*}
   {\bar x}^{ \infty} &= & (k_d- v+ c_\epsilon {\bar x}^{ \infty})(1-\delta ) + (k_u- v+ c_\epsilon {\bar x}^{ \infty})\delta=\frac{E [W]}{1-c_\epsilon}.
 \end{eqnarray*}}
Substituting ${\bar x}^{ \infty}$    we have the required bound:
$
     c_{\epsilon } < \frac{y- \overline{w}}{y- (1-\delta)(\overline{w}-\underline{w})}.
$
\eop

\noindent \textbf{Proof of Theorem \ref{Theorem_financenetwork}.(a)}: By  Lemma \ref{Lemma_threshold of q_eps}, the mapping  $\epsilon \mapsto q_\epsilon$ is   monotone; there exists
 ${\bar \epsilon}$, with  ${\bar \epsilon}_1 < {\bar \epsilon} \le {\bar \epsilon}_2$,  such that 
$q_\epsilon = 1-\delta$ for all 
$\epsilon \le {\bar \epsilon}_2$ and equals $1$ for the rest; here ${\bar \epsilon}_1$, ${\bar \epsilon}_2$   are given by Lemma \ref{Lemma_mono_PD}.


\noindent  {\bf Proof of part (b)} From  Lemma \ref{Lemma_threshold of q_eps}, 
${\bar \epsilon} < 1$ 
if and only if the conditions of part (b) are satisfied. 

\noindent  {\bf Proof of part (c)} By  the proof of Lemma \ref{Lemma_threshold of q_eps},  $g(\epsilon)$ definition, ${\bar \epsilon}$ equals the zero of the following (when condition \eqref{Eqn_cond_bareps}  is satisfied)
$\epsilon^2 + m_1 \epsilon + m_2 = 0$,  with, $m_1,m_2$ be  an appropriate constants.\eop
 \begin{lemma}
\label{Lemma_threshold of q_eps}
There exits a unique  ${\bar \epsilon}$, 
with ${\bar \epsilon}_1 \le {\bar \epsilon} \le {\bar \epsilon}_2$ and  with $\{{\bar \epsilon}_i\}$ as  in Lemma \ref{Lemma_mono_PD}, such that the   $q_\epsilon:=E[R^1 \ge R^2]$ satisfies the following threshold property:
\begin{equation}
    q_\epsilon =
    \begin{cases}
       1-\delta  & \text{ if }\ \epsilon < \bar{\epsilon} \\ 
      1 & \text{ if }\ \epsilon \ge  \bar{\epsilon}.
    \end{cases}
  \end{equation}
  Further ${\bar \epsilon} < 1$ if and only if equation \eqref{Eqn_cond_bareps} is  satisfied.
  \end{lemma}
  \noindent \textbf{Proof:}  
\noindent \textbf{Case A:} When $P_d = 0$ for some $\epsilon$,  the  returns of the agents are given by the following (recall $r_b \ge  r_s > d$, $w(1+d) \ge v$):

\vspace{-4mm}
{\small
\begin{eqnarray}
\label{eqn_R1eps_pd=0}
R^1 (\epsilon)  &=&  \left (  w\epsilon(1+r_s)+ \frac{(1-\alpha)(1-\epsilon)}{(\alpha +\epsilon)}  y  -v  \right )^+  \\
& & \hspace{-12mm} =\  \left (  w\epsilon(1+r_s)+  w (1-\epsilon) (1+r_b)  -v  \right )^+ \nonumber   = \  \left (    w (1+r_b) - v  + w \epsilon (r_s - r_b)   \right )  
 \mbox{,    and }  \\
R^2 (\epsilon) &=&    \left (K_i+\frac{\alpha (1+\epsilon) }{\alpha + \epsilon}   y -v-y \right )^+ 
=    \left ( K_i  -  w \epsilon (1+r_b)  -v  \right )^+  \label{eqn_R2eps_pd=0}  \\
& = & \left \{  
\begin{array}{llll}
R^2_u 
  &  \mbox{ w.p. }  \delta  &  \mbox{ where  }  R^2_u  := w (1+u) - v + w \epsilon (u-r_b)   \\
\left (  R^2_d   \right )^+ &  \mbox{ w.p. } 1-  \delta  &   \mbox{ where  }  R^2_d  :=    w (1+d) - v + w \epsilon (d-r_b)   . 
\end{array} \right .  \nonumber
\end{eqnarray}}
Hence with upward movement,  
$R^2 - R^1 = 
 R^2_u  -  R^1  
 =  w(u-r_b) +w\epsilon(u- r_s)  >  0. 
 $
 It is also  clear that   $ \left ( R^2_d  (\epsilon) \right )^+-  R^1 (\epsilon) < 0$.
Thus $q_\epsilon = 1-\delta$, when $P_d = 0$. By  Lemma \ref{Lemma_mono_PD}, this regime lasts for all $\epsilon$ satisfying  $0 \le \epsilon \le \bar{\epsilon}_1  $.
 
\noindent \textbf{Case B:} When $P_d = 1-\delta$, by Lemma \ref{Lemma_mono_PD},
we have ${\bar \epsilon}_1 < \epsilon \le {\bar \epsilon}_2$. In this case we have $ R^2_u >0$, but $R^1 (\epsilon) $ can be positive or zero.
 Since $R^2_d = 0$, we have that $q_\epsilon = P(R_1(\epsilon) \ge  R_2(\epsilon) ) \ge 1-\delta$. 
Further in this case,
 \begin{eqnarray*}
\label{eqn_R1eps}
R^1 (\epsilon)  &=&  \left (  w\epsilon(1+r_s)+ \frac{(1-\alpha)(1-\epsilon)}{(\alpha +\epsilon)}  {\bar x}^\infty  -v  \right )^+  \mbox{, and, } \\
R^2_u  &=&    \left (w(1+\epsilon)(1+u) +\frac{\alpha (1+\epsilon) }{\alpha + \epsilon}   {\bar x}^\infty -v-y \right ). 
 \end{eqnarray*}
 And hence
 \begin{eqnarray}
 \label{Eqn_diff_bewteen_returns}
 R^2_u  (\epsilon) -  R^1 (\epsilon)   & \le & w(1+u) +w\epsilon(u- r_s)-y +  {\bar x}^\infty\bigg(\frac{2\alpha +\epsilon -1}{\alpha +\epsilon}\bigg) \nonumber \\
 &=&    w(1+u) +w\epsilon(u- r_s) +  ({\bar x}^\infty -y) -{\bar x}^\infty \bigg( \frac{1-\alpha}{\alpha +\epsilon}\bigg) \nonumber \\
  &=&   w(u-r_b) +w\epsilon(u- r_s) +  ({\bar x}^\infty -y)\bigg(1- \frac{1-\alpha}{\alpha +\epsilon}\bigg) .
 \end{eqnarray}
If the upper bound on RHS  of \eqref{Eqn_diff_bewteen_returns} is negative
 then clearly $R^2_u (\epsilon) <  R^1(\epsilon)$.
 When the RHS is positive and $R^1(\epsilon) = 0$, then clearly  $R^2_u (\epsilon) >  R^1(\epsilon)$. 
 On the other hand, when RHS is positive and $R^1(\epsilon) > 0$, then the RHS is the exact value and not the  upper bound, and hence again $R^2_u (\epsilon) >  R^1(\epsilon)$.  Thus in all, $R^2_u (\epsilon) >  R^1(\epsilon)$ if and only if the RHS of \eqref{Eqn_diff_bewteen_returns} is positive.

 Hence with $P_d = 1-\delta$, we compute the following  and derive the required analysis by checking the negative/positive sign of the RHS of \eqref{Eqn_diff_bewteen_returns}. First observe from Lemma \ref{Lem_Average_clearing} that
 \begin{align}
\label{Eqn_difference}
({\bar x}^\infty - y )\bigg(1- \frac{1-\alpha}{\alpha +\epsilon}\bigg)   
= (1-\delta) \frac{ w(1+d) -v - w\epsilon(r_b-d)   }{\epsilon + \alpha \delta - \alpha \epsilon (1-\delta) 
  } \bigg(2\alpha+\epsilon-  1  \bigg)
\end{align}
  and consider the following function constructed (basically the denominator in the above is positive and multiply it with remaining terms of the RHS) using the RHS of  \eqref{Eqn_diff_bewteen_returns}:

  \vspace{-4mm}
  {\small
 \begin{eqnarray}
 \label{Eqn_bar_eps}
 g(\epsilon)& :=& w(u-r_b)(\epsilon+\alpha \delta -\alpha\epsilon(1-\delta))+w\epsilon(u- r_s) (\epsilon+\alpha \delta -\alpha\epsilon(1-\delta)) \nonumber\\
 &&+  (2\alpha -1+\epsilon)(1-\delta)(w(1+d) -v - w\epsilon(r_b-d)),\\
  g^{'}(\epsilon)&=&   w(u-r_b) (1-\alpha(1-\delta))+w(u-r_s)(2\epsilon+\alpha\delta-2\alpha\epsilon(1-\delta)) \nonumber\\
  && +(1-\delta)\bigg(w(1+d) -v - w\epsilon(r_b-d) -w(r_b-d) (2\alpha -1+\epsilon)\bigg), \nonumber\\
  g^{''}&=& 2w(u-r_s)(1-\alpha(1-\delta))-2w(1-\delta)(r_b-d),\nonumber\\
   g(\bar{\epsilon}_1)& :=& w(u-r_b)(\bar{\epsilon}_1+\alpha \delta -\alpha\bar{\epsilon}_1(1-\delta))+w\bar{\epsilon}_1(u- r_s) (\bar{\epsilon}_1+\alpha \delta -\alpha\bar{\epsilon}_1(1-\delta)) > 0 , \nonumber\\
  g(1)&=& w(1+2\alpha\delta- \alpha)(2u-r_s-r_b)+ 2\alpha(1-\delta)(w(1+d)-v-w(r_b-d)) \nonumber.
  \end{eqnarray}}
Observe that $g(\cdot)$ is concave in $\epsilon$  if  $(u-r_s)(1-\alpha(1-\delta))< (1-\delta)(r_b-d)$, else it is convex in $\epsilon$.
  
\textbf{Sub-case 1:} Consider the regime when $(u-r_s)(1-\alpha(1-\delta))< (1-\delta)(r_b-d)$. Observe that   $g(\bar{\epsilon}_1) > 0$ (by definition of ${\bar \epsilon}_1$). By concavity of $g(\cdot)$, the function (and hence the RHS of\eqref{Eqn_diff_bewteen_returns}) can at maximum change the sign one time. 
Further when the RHS is zero, clearly $q_\epsilon = 1$.  In other words there exists ${\bar \epsilon}$ with  $\bar{\epsilon}_1 \le {\bar \epsilon} \le \bar{\epsilon}_2$, such that $q_\epsilon = 1-\delta$ for all $\epsilon < {\bar \epsilon}$ and  equals $1$ 
when $ {\bar \epsilon} \le \epsilon < {\bar \epsilon}_2$.


\textbf{Sub-case 2:} Consider the regime  $(u-r_s)(1-\alpha(1-\delta)) \ge (1-\delta)(r_b-d)$. 
With this, it is easy to verify that
   $g(1) > 0$. 
%
  With $g(1) >0$ then we have $g^{'}(\bar{\epsilon}_1) > 0$, because  $2w(u-r_s)(1-\alpha(1-\delta))+ (1-\alpha)(1-2\alpha(1-\delta)) \ge 0$.  
  Once again, $g(\bar{\epsilon}_1) >0$
  In this case due  to convexity, $g(\epsilon)$ do not  changes sign   for all $\bar{\epsilon}_1 \le \epsilon \le \bar{\epsilon}_2$, and hence $q_\epsilon = 1-\delta$ for all $\bar{\epsilon}_1 \le \epsilon \le \bar{\epsilon}_2$.  

\noindent \textbf{Case C:} When $P_d = 1$, then $R^2 = 0$ a.s., while 
$R^1 \ge 0$, thus $q_\epsilon = 1$ and  this is for all  $\epsilon > \bar{\epsilon}_2$. Thus the threshold property.
Further one can have  ${\bar \epsilon} <  1$, either when   $g(1) < 0$ (which also ensures ${\bar \epsilon}_1 < 1$) or when ${\bar \epsilon}_2 < 1$  and hence the result.
\eop

 \begin{lemma}
\label{Lemma_mono_PD}
Define ${\bar \epsilon}_1 := \frac{w(1+d)-v}{w(r_b-d)} $. 
There exists ${\bar \epsilon}_2 \ge  {\bar \epsilon}_1$,   which is strictly greater when $ {\bar \epsilon}_1<1$ such that 
\begin{eqnarray}
P_d &=& \left \{
\begin{array}{lll}
    0    & \mbox{ if }  0 \le \epsilon \le {\bar \epsilon}_1 \\
    1-\delta
     &  \mbox{ if }  {\bar \epsilon}_1 < \epsilon \le {\bar \epsilon}_2, \\
        1
     &  \mbox{ if }   \epsilon > {\bar \epsilon}_2.
\end{array}
\right .
\label{Eqn_mono_Pd}
\end{eqnarray} 
Further ${\bar \epsilon}_2 < 1$ if and only if the second line 
of equation \eqref{Eqn_cond_bareps}  is satisfied. 
\ignore{

\vspace{-2mm}
{\small
\begin{eqnarray*}
\hspace{-2.5mm}
(1-\delta) (u-d) > \max \bigg \lbrace \frac{1+\alpha}{4w\alpha}\bigg(w(1+u)-v +w(u-r_b)\bigg)  ,\frac{u-r_b}{\alpha}\bigg \rbrace .
\end{eqnarray*}}}
\end{lemma}
{\bf Proof:}
First consider the following function of $\epsilon$, constructed using the bound $a_1$ and $c_\epsilon$  (the denominators  are  positive and then by making the denominator common in term $ c_\epsilon - a_1$) of Lemma \ref{Lem_Average_clearing}:

\vspace{-4mm}
{\small
\begin{eqnarray*}
h(\epsilon) = \bigg(\alpha +\alpha\epsilon\bigg)y- \bigg(y-k_d+v\bigg)\bigg(\alpha+\epsilon\bigg) 
  =  (\alpha+\epsilon) 
 \bigg  (w(1+d)   - v  + w \epsilon (d-r_b) \bigg ).
\end{eqnarray*}}
From Lemma \ref{Lem_Average_clearing}, if  $h(\epsilon) \ge 0$ then $P_d= 0$ and clearly this regime  lasts for all $\epsilon$, with  $0 \le \epsilon \le \bar{\epsilon}_1$. Further more,  beyond $\bar{\epsilon}_1$, $P_d > 0.$


Next
consider the following function of $\epsilon$, constructed similarly using the bound $a_2$ and $c_\epsilon$ of Lemma \ref{Lem_Average_clearing}:

\vspace{-4mm}
{\small
\begin{eqnarray}
 \label{Eqn_function_def}
    f(\epsilon) &=& \bigg(\alpha +\alpha\epsilon\bigg)\bigg(y-(1-\delta)w(1+\epsilon)(u-d)\bigg) - \bigg(y-k_u+v\bigg) \bigg(\alpha+\epsilon\bigg)\nonumber \\
    && \hspace{-16mm}=\ (\alpha +\epsilon) \bigg(w(1+d)-v -w\epsilon(r_b-d)\bigg)  +  \bigg ( (\alpha +\alpha \epsilon) \delta + (1-\alpha)  \epsilon \bigg )  w (u-d) (1+\epsilon). \hspace{2mm} 
 \end{eqnarray}}
 From  Lemma \ref{Lem_Average_clearing}, if  $f(\epsilon) \ge 0$ then  $ P_d \le 1-\delta$ 
 and if $f(\epsilon) < 0$  then  $ P_d =1$. 
 Therefore it  suffices to study this function. 
 By \eqref{Eqn_function_def}, $f(0)$, $f(1)$ and its  derivatives  are given by:

 \vspace{-4mm}
 {\small
 \begin{eqnarray*}
f(0) &=&  \alpha \bigg(w(1+d)-v +w\delta(u-d)\bigg) > 0,\\
  %
f(1) 
%
%
 &=& (1+\alpha) \bigg ( (w(1+d)-v)  - w(r_b -d) \bigg )  + w \bigg ( 4 \alpha \delta + 2(1-\alpha)  \bigg )
(u-d), \\
 f^{''}  (\epsilon) &=&  2w\bigg(u-r_b -\alpha(u-d)(1-\delta)\bigg)   \mbox{ for all $\epsilon$  $\bigg[$as $d-r_b = u-r_b - (u-d)\bigg]$, and, } \\
f^{'}(0) &=& w\bigg( \alpha (u-   r_b)   -(2 \alpha (1-\delta) -1) (u-d)\bigg)+ w(1+d) -v.
\end{eqnarray*}}

It is clear that   $f(0) >0$.  When  the second derivative $f^{''} <0$, the function is concave in $\epsilon$, then clearly
$f(\epsilon) \ge 0$ for all $\epsilon \le {\bar \epsilon}_2$ (for some $0< {\bar \epsilon}_2\le 1$) and $f(\epsilon) <0$ for the rest, where  $ {\bar \epsilon}_2 < 1$ if and only if $f(1) < 0$.


Now consider the case with $f^{''} \ge 0$, and then $f$ is convex (or linear). Under this condition,  clearly  the first derivative $f^{'}(0) > 0$ (as $\alpha^2 -2 \alpha (1-\delta) +1) >0$). Thus, $f(0) >0$ implies $f(\epsilon) > 0$ for all $\epsilon  \le {\bar \epsilon}_2$ and  so $ P_d \le (1-\delta)$ for all $\epsilon \le {\bar \epsilon}_2$, where ${\bar \epsilon}_2=1$. 

In all, we have the existence of an ${\bar \epsilon}_2$  such that $f(\epsilon) \ge 0$
(and hence 
$P_d \le 1-\delta$) if and only if $\epsilon \le {\bar \epsilon}_2$. Also  observe    from the second equality of \eqref{Eqn_function_def} that,   at $\epsilon= \bar{\epsilon}_1$, we have   $f(\bar{\epsilon}_1)> 0$. Thus  ${\bar \epsilon}_1 < {\bar \epsilon}_2$, whenever ${\bar \epsilon}_1 <1$. Recall $P_d = 0$ if and only if $\epsilon \le {\bar \epsilon}_1$.   Hence we have \eqref{Eqn_mono_Pd}.

Further $\bar{\epsilon}_2 < 1$  if and only if  $f(1) < 0$ (and $f'' <0$) which is equivalent 
to the second line of \eqref{Eqn_cond_bareps}.  \eop

We first begin with some  definitions. 
\begin{definition}
\label{Def_Asymptotically stable}
 {\bf Asymptotically stable (Attractor):} A set $A$ is said to be Asymptotically stable in the sense of Lyapunov,  which we refer as  attractor, if there exist a neighbourhood (called domain of attraction) starting in which the ODE trajectory converges to $A$ as time progresses (e.g., \cite{kushner2003stochastic}).
\end{definition}
\begin{definition}
  {\bf Equicontinuous in extended sense (\cite{kushner2003stochastic})):} Suppose that for each $n$, $h_n(.)$ is on the $\mathcal{R}^r$- valued measurable function on $(-\infty,\infty)$ and $\lbrace h_n(0)\rbrace$ is bounded. Also suppose that for each $T$ and $\varepsilon> 0$ there is a $\kappa > 0$  such that
 $$
 \limsup_{n} \sup_{0 \le t-s \le \kappa, \  t \le T} 
  \| h_n(t)-h_n(s)  \|  \le \varepsilon.
 $$
 Then we say that $\lbrace h_n(.)\rbrace$ is  equicontinuous  in the extended sense. By \cite[Theorem 2.2]{kushner2003stochastic},
 there exists a sub sequence that converges to some
continuous limit function.
\end{definition}

\noindent \textbf{Proof of Theorem \ref{Thm_RandomC_conv}:} We  prove the result using   \cite[Theorem 2.2, pp. 131]{kushner2003stochastic},  as  ${\bar g}_\epsilon(\cdot)$ is only measurable.
Towards this, we first need to prove (a.s.) equicontinuity in the extended sense  of the following sequence of two-dimensional functions defined for each $n$ (for any $t \ge 0$):   
$$[\epsilon^n(t), \psi^n(t) ] := [\epsilon_n, \psi_n ] + \hspace{-3mm} \sum_{i=n}^{m(t_n+t)-1}\hspace{-2mm} \gamma_i Y_i, \mbox {  } m(t) := \max\left  \{ n: \sum_{k=0}^{n-1}  \gamma_k \le t \right \}.$$  
This proof goes through almost exactly as in the proof of   \cite[Theorem 2.1, pp. 127]{kushner2003stochastic} and we follow exactly the same pattern. We begin with discussing some initial steps: i) the random vector  $Y_t$ depends on $(\epsilon_t, \psi_t)$ and    $(W_{t+1}, \Nw_{t+1})$; ii) observe  $\sup_t E\lvert Y_{t} \rvert ^2 < \infty$, which is trivially true by law of large numbers (LLN);  iii) clearly $\epsilon_t \le 1$  and $\psi_t \le 1+{\bar \Nw}$ for all $t$ and all sample paths;   iv)  we   have $E[Y_t \lvert {\cal G}_t] = \bar{\bf g} (\epsilon_t, \psi_t)+[e_t, 0]$ 
where,  
  $$e_t := E\bigg[\bigg(\frac{1}{\psi_{t+1}}- \frac{1}{\psi_{t}} \bigg)\bigg(W_{t+1} -\Nw_{t+1}\epsilon_t \bigg) \Bigg \lvert\mathcal{G}_t \bigg], $$
and v) the projection term in \cite{kushner2003stochastic}, $Z_t  \equiv  0$.

We further require to handle the difference term $e_t$.
We will now show that the error term $e_t$ is converges to zero in the limit and continue with the rest of the proof thereafter. 
 Towards this, observe that
  $\lvert \psi_{t} - \Nw_{t}\rvert  
  \le \psi_{t}+ \bar{\Nw} $ where $\bar{\Nw}$ is such that  $P(\Nw \le \bar{\Nw}) = 1$. Thus from \eqref{Eqn_psi_update},
 \begin{eqnarray}\bigg \lvert \frac{1}{\psi_{t+1}}- \frac{1}{\psi_{t}} \bigg \rvert = 
 \left \lvert \frac{ \gamma_t \left ( \psi_t - \Nw_{t}\right )}{\psi_{t} \psi_{t+1}} \right \rvert \le \frac{\gamma_t}{\psi_{t+1}} \left ( 1+ \frac{\bar{\Nw}}{\psi_{t}  } \right). \label{Eqn_Estimate}
 \end{eqnarray}
 
Consider the (almost sure) sample paths in which $\psi_t \to \   E[\Nw]  $  by LLN, as one can rewrite:  
\begin{eqnarray}
\label{Eqn_withswitch_addition_rand}
 \psi_{t} =  \frac{ n(t)}{t+n_0} = \frac{n_0 +  \sum_{k=1}^t \Nw_k }{t+n_0} .
\end{eqnarray}
For such sample paths (i.e., almost surely),  $\psi_{t} \ge  \varepsilon $ for all $t$ for some appropriate $\varepsilon > 0$ and  hence using \eqref{Eqn_Estimate}

\vspace{-2mm}
{\small
\begin{eqnarray}
\label{Eqn_further_errorbounding}
\lvert e_t \rvert \le \frac{\bar{s} \gamma_t}{\varepsilon} E\bigg[\big\lvert W_{t+1} - \Nw_{t+1}\epsilon_t \big \rvert  \bigg \rvert \mathcal{G}_t \bigg] \le \frac{ (2E[\Nw] +E[\Sw]) \bar{s} }{\varepsilon}  \gamma_t    \mbox{   a.s., with } {\bar s } := 1+\frac{{\bar \Nw}}{\varepsilon}, \hspace{3mm}
\end{eqnarray}}
and recall $W_{t+1}  =   \xi_t + \Xi_1(t) - \Xi_2(t)$.

Thus $e_t \to 0$  and $\sum_t \lvert e_t \rvert \gamma_t < \infty$ a.s.    The  update equation (starting at $n$ and for any $t\ge
0$)    can be written as below (as in \cite{kushner2003stochastic}):

\vspace{-2mm}
{\small
\begin{eqnarray*}
\label{Eqn_epsilon_integtral}
    [ \epsilon^{n}(t), \psi^n(t)]  &=&  [\epsilon_n, \psi_n] + \int_{0}^{t} \bar{\bf g} (\epsilon^n(s), \psi^n(s))ds + S^{n}(t) +\rho^n(t)  +  \sum_{i=n}^{m(t+t_n)-1} \gamma_i e_i \\ %
  \mbox{with, }  
     \rho^n(t) &:=&   \sum_{i=n}^{m(t+t_n)-1} \gamma_i\bar{\bf g} (\epsilon_n,\psi_n) - \int_{0}^{t} \bar{\bf g} (\epsilon^n(s),\psi^n(s))ds, \mbox{ and } \\
    S^n(t) &:=&  \sum_{i=n}^{m(t+t_n)-1} \hspace{-3mm} \gamma_i \Delta S_i = S_{m(t+t_n)-1} - S_n,   \  \Delta S_n \ = \ Y_n- \bar{\bf g} (\epsilon_n, \psi_n) -e_n.
\end{eqnarray*}}
Observe in the above that 
$\epsilon^n(t_k) = \epsilon_k$ for any $k > n$, where $t_k := \sum_{i <k} \gamma_i$. 
By LLN $\psi_k \to E[\Nw]$ in  almost all sample paths  and 
the  idea is to show the equicontinuity of the functions $\{\epsilon^n (\cdot), \psi^n(\cdot) \}_n$   for those sample paths; this guarantees the existence of limit function (along a sub-sequence) as in \cite{kushner2003stochastic}, and proceeding further as in \cite{kushner2003stochastic} we can show that the limit function satisfies the ODE \eqref{Eqn_g_beta}. 

The arguments required to show the equicontinuity in extended sense  are  exactly as in  \cite{kushner2003stochastic}, because of the following:
(i) $\{S_n\}$ is a martingale and using   well known  Martingale inequality (as in \cite{kushner2003stochastic}),
$
   \lim_m P \big \lbrace \sup_{j\ge m}\lvert S_j-S_m \rvert \ge \mu  \big \rbrace =0,
$ for any $\mu$,   because  
 
  $$\sum_i \gamma_i^2 < \infty  \mbox{ and   } E[\Delta S_i \Delta S_j] =E[E[\Delta S_i \Delta S_j\rvert {\cal G}_j ]]  = 0, \forall j < i; $$ 
 (ii) recall  $\psi_k \to E[\Nw]$ as $k\to \infty$ for  chosen sample paths; there exists a $C_g, {\bar k}  < \infty$ such that
 $ \lvert{\bar g}_\epsilon(\epsilon_k, \psi_k) \rvert \le \lvert \beta \rvert  /\psi_k < C_g$,
 $\lvert E[\Nw] - \psi_k \rvert \le C_g$ and 
   hence
 $\sup_{t} \lvert \rho^k (t) \rvert  \le 2 C_g \gamma_k  $ for 
 all    $k\ge {\bar k}$;  (iii)    the sequence $\{[\epsilon^n(0),  \psi^n(0) ]\}_{n\ge \bar{k}}
= \{[\epsilon_n, \psi_n]\}_{n \ge \bar{k}}
$  is bounded a.s. by $[1, 1+{
\bar \Nw}]$;
  and    (iv) finally for any $t \ge  t'$: 
\begin{eqnarray*}
&& \bigg \lvert \int_{0}^{t} {\bar g}_\epsilon(\epsilon^n(s),\psi^n(s))ds -  \int_{0}^{t'} {\bar g}_\epsilon(\epsilon^n(s),\psi^n(s))ds \bigg \rvert  
 \\
 &\le &   (t-t') \sup_{ \epsilon \in [0,1], k \ge n} \lvert {\bar g}_\epsilon(\epsilon,\psi_k) \rvert = C_g (t-t') .
\end{eqnarray*}
Hence with $\Theta^n (\cdot) := (\epsilon^n(\cdot), \psi^n(\cdot))$, the sequence $\{ \Theta^n(\cdot) \}_n$ is equicontinuous in extended sense almost surely (observe again that the above proof is using similar arguments as in proof of \cite[Theorem 2.1]{kushner2003stochastic}, with extensions to measurable ${\bar g}$ made possible because of boundedness of ${\bar g}$).

 In  Corollary \ref{Lemma_ODE_Analysis} we have identified the attractors of   \eqref{Eqn_g_beta},  and showed that  the combined domain of attraction is the whole   $[0,1] \times [0, {\bar C}]$ (for any ${\bar C} <  \infty$). 
 Choose ${\bar C}$ such that 
   the dynamics visits  $[0,1]\times [0, {\bar C}]$ infinitely often (possible because $P({\cal N}_t < {\bar \Nw}) =1$)   and hence converges (a.s.) to  one of the limit points of Theorem  \ref{Theorem_financenetwork} and $\beta$ by \cite[Theorem 2.2, pp. 131]{kushner2003stochastic}. \eop

\noindent \textbf{Proof of Theorem \ref{Thm_RandomC_convwithdefault}:} 
We need to prove (a.s.) equicontinuity in the extended sense  of the  sequence $[\epsilon^n(t), \psi^n(t) ]$ of two-dimensional functions defined for each $n$  and 
the proof goes through almost exactly as in the proof of   Theorem \ref{Thm_RandomC_conv}.  We mention the modifications: i) the random vector  $Y_t$ depends on $(\epsilon_t, \psi_t)$ and    $(W_{t+1}, \Nw_{t+1}, \mathcal{L}_{t+1})$; ii) clearly $\epsilon_t \le 1$  and $\psi_t \le 1+{\bar \Nw} + {\bar {\cal L}}$ for all $t$ and all sample paths;   iii)  we   have $E[Y_t \lvert {\cal G}_t] = \bar{\bf g} (\epsilon_t, \psi_t)+[e_t, 0]$ 
where,  
  $$e_t := E\bigg[\bigg(\frac{1}{\psi_{t+1}}- \frac{1}{\psi_{t}} \bigg)\bigg(W_{t+1} -\Nw_{t+1}\epsilon_t + \mathcal{L}_{t+1} \epsilon_t \bigg) \Bigg \lvert\mathcal{G}_t \bigg], $$
and iv) the projection term in \cite{kushner2003stochastic}, $Z_t  \equiv  0$.

We will now show that the error term $e_t$ converges to zero. Towards this, first observe that
  $\lvert \psi_{t} - \Nw_{t} -\mathcal{L}_t \rvert  
  \le \psi_{t}+ \bar{\Nw} + {\bar {\cal L}} $ where $\bar{\Nw}$ and ${\bar {\cal L}}$ are such that  $P(\Nw \le \bar{\Nw}) = 1$ and   $P(\mathcal{L} \le {\bar {\cal L}}) = 1$. Thus from \eqref{Eqn_with_additionswitch_updaterulewithdefault},
 \begin{eqnarray}\bigg \lvert \frac{1}{\psi_{t+1}}- \frac{1}{\psi_{t}} \bigg \rvert = 
 \left \lvert \frac{ \gamma_t \left ( \psi_t - \Nw_{t} - \mathcal{L}_t \right )}{\psi_{t} \psi_{t+1}} \right \rvert \le \frac{\gamma_t}{\psi_{t+1}} \left ( 1+ \frac{\bar{\Nw} +{\bar {\cal L}}}{\psi_{t}  } \right). \label{Eqn_Estimatewithbounded}
 \end{eqnarray}

 Further, observe that 
$$
\psi_{t+1} \ge \psi_t + \frac{1}{t+n_0+1} \left ( \Nw_t -{\cal L}_t  - \psi_t \right),
$$hence the sequence $\{\psi_t\}$ can be lower bounded (term-wise) by a sequence of sample means constructed using  $\Nw_t - {\cal L}_t$ as in \eqref{Eqn_withswitch_addition_rand}
 and this ensures that $\psi_t \ge \varepsilon$  where $\varepsilon$ now depends upon $E[{ {\cal N}}] - E[{ {\cal L}}]$ which is  strictly positive by hypothesis. The final bound \eqref{Eqn_further_errorbounding} now changes with modified ${\bar s } = 1 + \nicefrac{(\bar{\Nw} + {\bar {\cal L}})}{\varepsilon  }$, and as $\Nw_{t+1}-{\cal D}_{t+1} < \Nw_{t+1}$:
 
\vspace{-2mm}
{\small
\begin{eqnarray*}
\lvert e_t \rvert \le \frac{\bar{s} \gamma_t}{\varepsilon} E\bigg[\big\lvert W_{t+1} - (\Nw_{t+1}-{\cal D}_{t+1})\epsilon_t \big \rvert  \bigg \rvert \mathcal{G}_t \bigg] \le \frac{ (2E[\Nw] +E[\Sw]) \bar{s} }{\varepsilon}  \gamma_t    \mbox{   a.s.,  }   
\end{eqnarray*}}and recall $W_{t+1}  =   \xi_t + \Xi_1(t) - \Xi_2(t)$.
Thus $e_t \to 0$  and $\sum_t \lvert e_t \rvert \gamma_t < \infty$ a.s. 
The rest of the proof follows in exactly the same lines as that in Theorem \ref{Thm_RandomC_conv}, now using the fact that $\psi_t $ for large $t$ can be lower bounded by $\varepsilon$ almost surely and using the modified bounds.

Thus with $\Theta^n (\cdot) := (\epsilon^n(\cdot), \psi^n(\cdot))$, the sequence $\{ \Theta^n(\cdot) \}_n$ is equicontinuous in extended sense almost surely (say on set ${F}$), by using remaining steps as in  the
proof of Theorem \ref{Thm_RandomC_conv}.
Hence by  extended version of Arzela-Ascoli Theorem \cite[section 4, Theorem 2.2, pp. 127]{kushner2003stochastic}, there exists a sub-sequence $(\Ups^{k_m}(\omega, \cdot))$ which converges to some continuous limit, call it $\ups(\omega, \cdot)$, uniformly on each bounded interval for $\omega \in F$ such that, with ${\bar {\bf g}}^D := ({\bar  g}_\epsilon^D, {\bar   g}_\psi^D)$, 
\begin{align}\label{eqn_ups_inf}
\ups(t) = \lim_{k_m \to \infty} \Ups_{k_m}(\omega) + \int_0^t {\bar {\bf g}}^D (\ups(s)) ds.
\end{align}
Thus, for every $\epsilon > 0$ and $T > 0$, there exists $N_\epsilon^T$ such that (note that ${\Ups}^{k_m}(t) = {\Ups}_l$ for $t = t_l - t_{k_m}$ ($l > k_m$) such that $0 \leq t \leq T$): 
\begin{align}\label{eqn_dist_scheme_ODE_}
\sup_{l: t_l \in [t_{k_m},   \ T + t_{k_m} ]} d({\Ups}_l , \ups(t_l - t_{k_m})) \leq \epsilon/2 \mbox{ for all } k_m \geq N_\epsilon^T.
\end{align}
   This completes part (i).
Part (ii) follows by equicontinuity and   by \cite[Theorem 2.2, pp. 131]{kushner2003stochastic} under assumption {\bf A}. 
\eop



\end{document}